\documentclass[final,5p,times,twocolumn]{elsarticle}
\usepackage{hyperref}
\usepackage{lineno,color}
\modulolinenumbers[5]
\journal{Astroparticle Physics}
\usepackage{amsmath}
\usepackage{graphicx}
\usepackage{epstopdf}
\newcommand\aap{A\&A}
\newcommand\apjl{ApJ}

\def\procspie{\ref@jnl{Proc.~SPIE}}   

\usepackage{soul}
\usepackage{float}
\usepackage{fixltx2e}

\biboptions{authoryear}

\begin{document}

\begin{frontmatter}

\title{\vspace{-2.2cm}Performance of the MAGIC telescopes under moonlight}

%

\author[a]{\vspace{-0.3cm}M.~L.~Ahnen}
\author[b,u]{S.~Ansoldi}
\author[c]{L.~A.~Antonelli}
\author[d]{C.~Arcaro}
\author[e]{A.~Babi\'c}
\author[f]{B.~Banerjee}
\author[g]{P.~Bangale}
\author[g,**]{U.~Barres de Almeida}
\author[h]{J.~A.~Barrio}
\author[i,j]{J.~Becerra Gonz\'alez}
% ...

\address[a]{ETH Zurich, CH-8093 Zurich, Switzerland}
\address[b]{Universit\`a di Udine, and INFN Trieste, I-33100 Udine, Italy}
\address[c]{INAF - National Institute for Astrophysics, viale del Parco Mellini, 84, I-00136 Rome, Italy}
\address[d]{Universit\`a di Padova and INFN, I-35131 Padova, Italy}
\address[e]{Croatian MAGIC Consortium, Rudjer Boskovic Institute, University of Rijeka, University of Split - FESB, University of Zagreb - FER, University of Osijek,Croatia}
\address[f]{Saha Institute of Nuclear Physics, 1/AF Bidhannagar, Salt Lake, Sector-1, Kolkata 700064, India}
\address[g]{Max-Planck-Institut f\"ur Physik, D-80805 M\"unchen, Germany}
\address[h]{Universidad Complutense, E-28040 Madrid, Spain}
\address[i]{Inst. de Astrof\'isica de Canarias, E-38200 La Laguna, Tenerife, Spain}
\address[j]{Universidad de La Laguna, Dpto. Astrof\'isica, E-38206 La Laguna, Tenerife, Spain}
% ...
\address[u]{Japanese MAGIC Consortium, ICRR, The University of Tokyo, Department of Physics and Hakubi Center, Kyoto University, Tokai University, The University of Tokushima, Japan}
% ...
\address[**]{now at Centro Brasileiro de Pesquisas F\'isicas (CBPF/MCTI), R. Dr. Xavier Sigaud, 150 - Urca, Rio de Janeiro - RJ, 22290-180, Brazil}
% ...
\cortext[mycorrespondingauthor]{Corresponding authors: Daniel Guberman (dguberman@ifae.es) and Pierre Colin (colin@mppmu.mpg.de)}

\begin{abstract}
MAGIC, a system of two imaging atmospheric Cherenkov telescopes, achieves its best performance under dark conditions, i.e.\ in absence of moonlight or twilight. Since operating the telescopes only during dark time would severely limit the duty cycle, observations are also performed when the Moon is present in the sky. Here we develop a dedicated Moon-adapted analysis to characterize the performance of MAGIC under moonlight. We evaluate energy threshold, angular resolution and sensitivity of MAGIC under different background light levels, based on Crab Nebula observations and tuned Monte Carlo simulations. This study includes observations taken under non-standard hardware configurations, such as reducing the camera photomultiplier tubes gain by a factor $\sim$1.7 (reduced HV settings) with respect to standard settings (nominal HV) or using UV-pass filters to strongly reduce the amount of moonlight reaching the cameras of the telescopes. The Crab Nebula spectrum is correctly reconstructed in all the studied illumination levels, that reach up to 30 times brighter than under dark conditions. The main effect of moonlight is an increase in the analysis energy threshold and in the systematic uncertainties on the flux normalization. The sensitivity degradation is constrained to be below 10\%, within 15-30\% and between 60 and 80\% for nominal HV, reduced HV and UV-pass filter observations, respectively. No worsening of the angular resolution was found. Thanks to observations during moonlight, the maximal duty cycle of MAGIC can be increased from $\sim$18\%, under dark nights only, to up to $\sim$40\% in total with only moderate performance degradation.
\end{abstract}


\end{frontmatter}


\section{Introduction}

In the last decades the Imaging Atmospheric Cherenkov Technique (IACT) opened a new astronomical window to observe the $\gamma$-ray sky at Very High Energy (VHE, E$>$50~GeV). After the pioneering instruments of the last century, the three most sensitive currently operating instruments, VERITAS \citep{VERITAS2008}, H.E.S.S.\citep{HESS2006} and MAGIC \citep{upgrade1}, have discovered more than a hundred sources, comprised of a large variety of astronomical objects (see \cite{DeNaurois2015} for a recent review). The IACT uses one or several optical telescopes that image the air showers induced by cosmic $\gamma$ rays in the atmosphere, through the Cherenkov radiation produced by the ultra-relativistic charged particles of the showers. The air-shower Cherenkov light peaks in the optical/near-UV band. This faint light flash can be detected above the ambient optical light background using fast photodetectors. The IACT works only by night and preferentially during dark moonless conditions.

IACT telescope arrays are usually optimized for dark nights, using as photodetectors UV-sensitive fast-responding photomultiplier tubes (PMTs), ideal to detect the nanosecond Cherenkov flash produced by an air shower. PMTs can age (gain degradation with time) quickly in a too bright environment, which restricts observations to relatively dark conditions. When IACT instruments operate only during moonless astronomical nights, their duty cycle is limited to 18\% ($\sim$1500\,h/year), without including the observation time loss due to bad weather or technical issues. Every month around the full Moon, the observations are generally fully stopped for several nights in a row.

Operating IACT telescopes during moonlight and twilight time would allow increasing the duty cycle up to $\sim$40\%. This is interesting for many science programs, to obtain a larger amount of data and a better time coverage without full-Moon breaks. It may also be crucial for the study of transient events (active galaxy nucleus flares, $\gamma$-ray bursts, cosmic neutrino or gravitational wave detection follow-ups, etc.) that occur during moonlight time. With moonlight observation, the IACT can be more reactive to the variable and unpredictable $\gamma$-ray sky. Moreover, operation under bright background light offers the possibility to observe very close to the Moon to study for instance the cosmic-ray Moon shadow to probe the antiproton and positron fractions \citep{ElectronMoonShadow,FirstMoonShadow} or the lunar  occultation of a bright $\gamma$-ray source, which was used e.g.~in hard X-ray for source morphology studies \citep{1975natur}.

Different hardware approaches have been developed by IACT experiments to extend their duty cycle into moonlight time. One possibility is to restrict the camera sensitivity to wavelengths below 350\,nm, where the moonlight is absorbed by the ozone layer. This idea was applied to the Whipple 10\,m telescope, which was equipped with the dedicated UV-sensitive camera ARTEMIS \citep{ArtemisFilter}, or with a simple UV-pass filter in front of the standard camera \citep{WhippleFilter}. The drawback of this technique is the dramatic increase of the energy threshold (a factor $\sim$4) due to the reduction of the collected Cherenkov light. The CLUE experiment \citep{CLUE} was a similar attempt with an array of 1.8\,m telescopes sensitive in the background-free UV range 190-230\,nm. More recently, the VERITAS collaboration also developed UV-pass filters to extend the operation during moonlight time \citep{VeritasFilter}. Another approach, developed first by the HEGRA collaboration \citep{HegraMoon}, is to reduce the High Voltage (HV) applied to the PMTs (reducing the gain) to limit the anode current that can damage the PMTs. This, however, only allows observations at large angular distances from a partially illuminated Moon. An alternative way to safely operate IACT arrays under moonlight would be to use, instead of PMTs, silicon photomultiplier detectors, which are robust devices that can be exposed to high illumination levels without risk of damages. This was successfully demonstrated with the FACT camera \citep{FactMoon}, which can operate with the full Moon inside its field of view (FOV). The use of a silicon photomultiplier camera is actually under consideration for the new generation of IACT instruments \citep{SiPM-SST1M,SiPM-SCT,SiPM-LST,SiPM-ASTRI,Light-trap}.

The cameras of the MAGIC telescopes, which are equipped with low-gain PMTs, were designed from the beginning to allow observations during moderate moonlight \citep{OldMoon, MagicMoon-ICRC2009}. The use of reduced HV \citep{MagicMoonShadow} and UV-pass filters \citep{Filters} were introduced later to extend the observations to all the possible Night Sky Background (NSB) levels, up to few degrees from a full Moon.

IACT observations under moonlight are becoming more and more standard, and
are routinely performed with the MAGIC and VERITAS telescopes. The
performance of VERITAS under moonlight with different hardware settings at
a given NSB level has been recently reported \citep{VeritasNew}. In this
paper, we present a more complete study on how the performance of an IACT
instrument is affected by moonlight and how it degrades as a function of
the NSB. Our study is based on extensive observations of the Crab Nebula,
adapted data reduction and tuned Monte Carlo (MC) simulations. The
observations, carried out from October 2013 to March 2016 by MAGIC with
nominal HV, reduced HV and UV-pass filters, cover the full range of NSB
levels that are typically encountered during moonlight nights.

\section{The MAGIC telescopes under moonlight} \label{sec:MAGIC}

MAGIC (Major Atmospheric Gamma-ray Imaging Cherenkov) is a system of two 17\,m-diameter imaging atmospheric Cherenkov telescopes located at the Roque de los Muchachos Observatory on the Canary Island of La Palma, Spain, at an altitude of 2200\,m a.s.l. The telescopes achieve their best performance for VHE $\gamma$-ray observation in the absence of moonlight. Under such conditions, and for zenith angles below $30^\circ$, MAGIC reaches an energy threshold of $\sim50$\,GeV at trigger level, and a sensitivity above 220\,GeV of $0.67 \pm 0.04 \% $ of the Crab Nebula flux (Crab Units, C.U.) in 50 hours of observation \citep{upgrade2}.

MAGIC is also designed to observe under low and moderate moonlight. Each camera consists of 1039 6-dynode PMTs, that are operated at a relatively low gain, typically of 3-4~$ \times 10^4$. This configuration was set specifically to decrease the amount of charge that hits the last PMT dynode (anode) during bright sky observations due to the Moon, preventing fast aging (see more details in Section 3.10 of \cite{upgrade1}). With the same criteria, there are established safety limits for the current generated in the PMTs. Individual pixels (PMT) are automatically switched off if their anode currents (DCs) are higher than 47\,$\mu$A and the telescopes are typically not operated if the median current in one of the cameras is above 15\,$\mu$A  (as a reference, during dark time the median current is about 1\,$\mu$A). A detailed study on the gain drop of the MAGIC PMTs when exposed to high illumination levels was reported in \cite{OldMoon}, which shows that while the detectors are operated at low gain and within the imposed safety limits no significant degradation is expected in the lifetime of MAGIC.

\subsection{The MAGIC trigger system}\label{sec:trigger}
The standard MAGIC trigger has three levels. The first one (L0) is an amplitude discriminator that operates individually on every pixel of the camera trigger area. All the L0 signals are sent to the second level (L1), a digital system that operates independently on each telescope, looking for time-coincident L0 triggers in a minimum number of neighboring pixels (typically three). Finally, the third level (L3) looks for time coincidence of the L1 triggers of each telescope \citep{upgrade1}.

The trigger rates depend on the discriminator threshold (DT) set on each PMT at the L0 level. The DTs are controlled by the Individual Pixel Rate Control (IPRC) software, which aims to keep stable the L0 rates of every pixel within certain desired limits. These limits are optimized to provide the lowest possible energy threshold while keeping accidental rates at a low level which can be handled by the data acquisition system (DAQ) without incurring a significant additional dead time. The accidental L0 triggers are dominated by NSB fluctuations. As they can vary significantly during observations, the DTs are constantly changed by the IPRC. If the L0 rate of one pixel moves temporary outside the imposed limits, as it could happen if, e.g., a bright star is in the FOV, the IPRC adjusts its DT until the rate is back within the desired levels (for more details see Section 5.3.4 of \cite{upgrade1}). Noise fluctuations are higher in a region with high density of bright stars, like the galactic plane, than in an extragalactic one. During relatively bright moonlight observations the main contribution to NSB comes from the Moon itself. Unlike stars, that only affect a few pixels, the moonlight scattered by the atmosphere affects the whole camera almost uniformly (with the exception of the region within a few degrees of the Moon). The induced noise depends on zenith angle, the angular distance between the pointing direction and the Moon, its phase, its position in the sky and its distance to the Earth \citep{Britzger}. Essentially, accidental L0 rates get higher during moonlight observations and IPRC reacts increasing the DTs, resulting in a higher trigger-level energy threshold.

\subsection{Moonlight observations}

In this work, the performance of MAGIC is studied for different NSB conditions. During the observations we do not measure directly the NSB spectrum, but just monitor the DC in every camera pixel. We infer the NSB level by comparing the measured median DC in the camera of one of the telescopes, MAGIC~1, with a reference average median DC that is obtained in a well-defined set of observation conditions. Here we use as reference the telescopes pointing toward the Crab Nebula at low zenith angle during astronomical night, with no Moon in the sky or near the horizon, and good weather (no clouds or dust layer). We shall refer to these conditions as $\textit{NSB}_{\text{Dark}}$. The median DC in MAGIC~1 during Crab dark observations is affected by hardware interventions: it depends on the PMTs HV and as so it can change after a camera flat-fielding. For the whole studied period Crab median DC during dark observations with nominal HV lies between 1.1 and 1.3\,$\mu$A\footnote{As the Crab Nebula is in the galactic plane, the NSB is lower by 30-40\% for a large fraction of MAGIC observations, which point to extragalactic regions of the sky. During reduced HV and UV-pass filter observations the measured DC is lower than what would be obtained if observing under the same NSB conditions and nominal HV. Correction factors are applied to properly convert from DC to NSB level based on the gain reduction factor of the PMTs and on the moonlight transmission of the filters.}. 

Due to the constraints imposed by the DC safety limits described in Section \ref{sec:trigger}, observations are possible up to a brightness of about 12\,$\times \,\textit{NSB}_{\text{Dark}}$ using the standard HV settings (nominal HV). Observations can be extended up to about 20\,$\times  \, \textit{NSB}_{\text{Dark}}$ by reducing the gain of the PMTs by a factor $\sim$1.7 (reduced HV settings). When the HV is reduced there is less amplification in the dynodes and so fewer electrons hit the anode. However, the PMT gains cannot be reduced by an arbitrary large factor because the performance would significantly degrade, resulting in lower collection efficiency\footnote{In MAGIC the HV divider chain is fixed for all dynodes and the voltage is also reduced at the first dynode.}, slower time response, larger pulse-to-pulse gain fluctuations and an intrinsically worse signal-to-noise ratio \citep{Photonis}.

Even when the telescopes are operated with reduced HV, observations are severely limited or cannot be performed if the Moon phase is above $90\%$. Observations can, however, be extended up to about 100\,$\times  \, \textit{NSB}_{\text{Dark}}$ with the use of UV-pass filters. This limit is achievable if the filters are installed and at the same time PMTs are operated with reduced HV. This is done only in extreme situations ($>$50\,$\times  \, \textit{NSB}_{\text{Dark}}$). All the UV-pass filter data included in this work were taken with nominal PMT gain. In practice, observations can be performed in conditions that are safe for the PMTs as close as a few degrees away from a full Moon. The telescopes can be pointed almost at any position in the sky, regardless the Moon phase, and, as a result, they can be operated continuously without full Moon breaks \citep{Filters}. The characteristics of the filters are explained in Section \ref{sec:filters}.

\begin{figure}[t]
\includegraphics[width=\columnwidth]{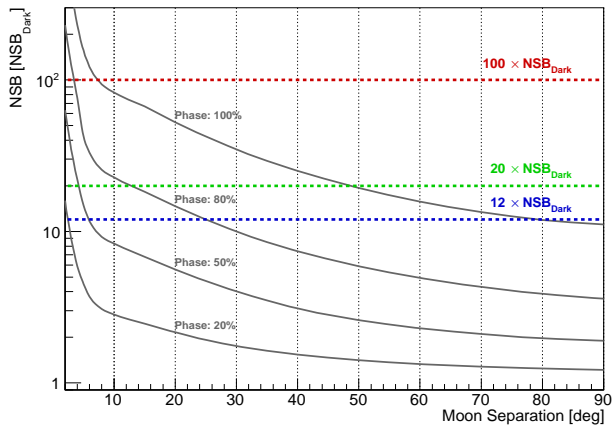}
\caption{Crab FOV brightness, simulated with the code described in \cite{Britzger}, as a function of the angular distance to the Moon for different Moon phases (gray solid lines). Moon zenith angle was fixed at 45$^\circ$. In blue, green and red the maximum NSB levels that can be reached using nominal HV, reduced HV and UV-pass filters are shown, respectively.}\label{fig:NSBvsDist}
\end{figure}

As a first approximation, the brightness of the whole sky strongly depends on the Moon phase and its zenith angle. Figure \ref{fig:NSBvsDist} shows the brightness of a Crab-like FOV, seen by MAGIC, as a function of the angular distance to the Moon for different Moon phases. The brightness values were simulated with the code described in \cite{Britzger}, for a Moon zenith angle of 45$^\circ$. While the Moon phase is lower than $50\%$, the brightness is below 5\,$\times  \, \textit{NSB}_{\text{Dark}}$ in at least $80\%$ of the visible sky and then in general operations can be safely performed with nominal HV. For phases larger than $80\%$, the brightness is typically above 10\,$\times  \, \textit{NSB}_{\text{Dark}}$ in most of the sky when the Moon is well above the horizon, and the observations are usually only possible with reduced HV. When the Moon phase is close to 100\%, observations are practically impossible without the use of UV-pass filters. Combining nominal HV, reduced HV and UV-pass filter observations, MAGIC could increase its duty cycle to $\sim$40\%. 

\subsection{UV-pass filters}\label{sec:filters}

\begin{figure}[t]
\includegraphics[width=\columnwidth]{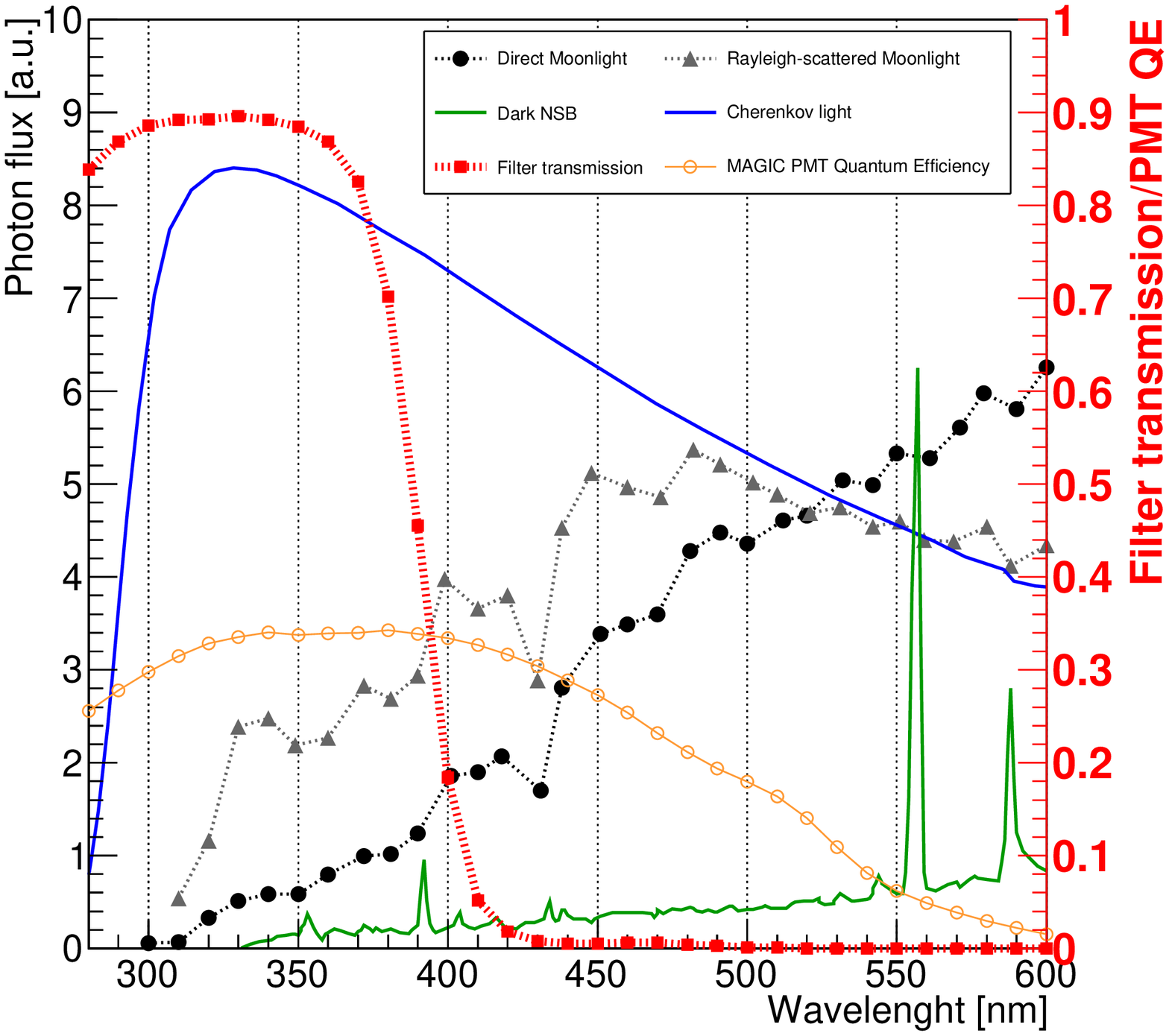}
\caption{The blue curve shows the typical Cherenkov light spectrum for a vertical shower initiated by a 1\,TeV $\gamma$ ray, detected at 2200\,m a.s.l \citep{1TeVshower}. In green, the emission spectrum of the NSB in the absence of moonlight measured in La Palma \citep{Benn98}. The dotted curves show the shape of direct moonlight spectrum (black) and Reyleigh-scattered diffuse moonlight (grey) \citep{SMARTS1,SMARTS2}. The four curves are scaled by arbitrary normalization factors. The filter transmission curve is plotted in red. As a reference, the quantum efficiency of a MAGIC PMT is plotted in orange (using the right-hand axis).}\label{fig:filters_transmission}
\end{figure}

\begin{figure}[t]
\includegraphics[width=\columnwidth]{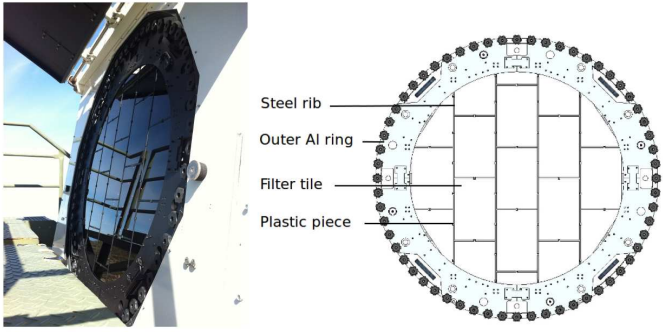}
\caption{On the left, the UV-pass filters installed on the camera of one of the MAGIC telescopes. On the right, design of the frame that holds the filters. The outer aluminium ring is screwed to the camera.}\label{fig:frame}
\end{figure}

Camera filters are used to reduce strongly the NSB light, while preserving a large fraction of the Cherenkov radiation that peaks at $\sim$330\,nm. The filter transmission must be high in UV and cut the longer wavelengths. 
They were selected to maximize the signal-to-noise ratio that scales as $T_{\text{Cher}}/\sqrt{T_{\text{Moon}}}$, with $T_{\text{Cher}}$ and $T_{\text{Moon}}$ the Cherenkov-light and the moonlight transmission of the filters, respectively.
An additional constraint was imposed by the MAGIC calibration laser, which has a wavelength of 355\,nm. $T_{\text{Moon}}$ depends on the spectral shape of the scattered moonlight, which depends on the angular distance to the Moon. Far from it (tens of degrees away) the NSB is dominated by Rayleigh-scattered moonlight that peaks at $\sim$470\,nm. Close to the Moon, Mie scattering of moonlight dominates; its spectrum peaks at higher wavelengths and resembles more the spectrum of the light coming directly from the Moon (``direct moonlight''). The spectral shape of the NSB is also affected by the aerosol content and distribution, and by the zenith angle of the Moon.
 
Typical spectra for Rayleigh-scattered and direct moonlight were computed using the code SMARTS \citep{SMARTS1,SMARTS2}, adding the effect of the Moon albedo. They can be seen in Figure~\ref{fig:filters_transmission}, together with the spectrum of the Cherenkov light from a vertical shower initiated by a 1\,TeV $\gamma$ ray, at 2200\,m a.s.l.~\citep{1TeVshower}. Taking the spectral information of Cherenkov light and diffuse moonlight into account, we selected commercial inexpensive UV-pass filters produced by Subei\footnote{{http://www.globalsources.com/sbgx.co}} (model ZWB3) with a thickness of 3\,mm and a wavelength cut at 420\,nm. The filter transmission curve is also shown in Figure~\ref{fig:filters_transmission}. The transmission of the filters for Cherenkov light from air showers were measured by installing a filter in only one of the two telescopes, selecting image of showers with similar impact parameters (defined as the distance of the shower axis to the telescope center) for both telescopes, and comparing the integrated charge in both images. The measured Cherenkov-light transmission at 30$^\circ$ from zenith is $T_{\text{Cher}}=(47 \pm 5)\%$ . The transmission for the NSB goes from $\sim$20\%, when pointing close to the Moon, to $\sim$33\%, when background light is dominated by either Rayleigh-scattered moonlight or the dark NSB. Other parameters such as the Moon phase and zenith angle also affect the NSB transmission. The conversion from DC to NSB level could then be different depending on the observation conditions. For the performance study in this work we adopted a ``mean scenario'', corresponding to an NSB transmission of 25\%. 

The filters were bought in tiles of 20\,cm\,$\times$\,30\,cm, and mounted on a light-weight frame. This frame consists of an outer aluminum ring that is screwed to the camera and steel  6\,mm\,$\times$\,6\,mm section ribs that are placed between the filter tiles (see Figure~\ref{fig:frame}). The filter tiles are fixed to the ribs by plastic pieces and the space between tiles and ribs is filled with silicon. This gives mechanical stability to the system and prevents light leaks. Two people can mount, or dismount, the UV-pass filter on a MAGIC camera in about 15 minutes.

\section{Data sample and analysis methods}\label{sec:DatAna}

To characterize the performance of MAGIC under moonlight we used 174 hours of Crab Nebula observations taken between October 2013 and January 2016, under NSB conditions going from 1 (dark) up to $30 \times  \, \textit{NSB}_{\text{Dark}}$\footnote{Observations are possible at higher illumination levels, but it is hard to get Crab data under such occasions. In fact, only on rare situations MAGIC targets are found under higher NSB levels than the ones analyzed in this work.}. Observations were carried out in the so-called wobble mode \citep{Fomin}, with a standard wobble offset of $0.4^\circ$. All the data correspond to zenith angles between 5$^\circ$ and 50$^\circ$. For this study we selected samples that were recorded during clear nights, for which the application of the MC corrections described in \citep{fruck2013} are not required.


Data were divided into different samples according to their NSB level and the hardware settings in which observations were performed (nominal HV, reduced HV or UV-pass filters), as summarized in Table~\ref{tabTime}. When dividing the data we aimed to have rather narrow NSB bins while keeping sufficient statistics in each of them ($\sim10$ hours per bin). Bins are slightly wider in the case of the UV-pass filter data to fulfill that requirement.  

 \begin{table}[t]
\centering
\begin{tabular}{| c | c | c |}
\hline
Sky Brightness & Hardware Settings & Time \\

[$\textit{NSB}_{\text{Dark}}$]  &  &  [h]  \\
\hline
   \hline

     1 (Dark) & nominal HV & 53.5 \\
     1-2 & nominal HV & 18.9 \\     
     2-3 & nominal HV & 13.2 \\     
     3-5 & nominal HV & 17.0 \\     
     5-8 & nominal HV & 9.8 \\     
        \hline
        
     5-8 & reduced HV & 10.8 \\     
     8-12 & reduced HV & 13.3 \\
     12-18 & reduced HV & 19.4 \\     
        \hline
        
     8-15 & UV-pass filters & 9.5 \\
     15-30 & UV-pass filters & 8.3 \\     
   \hline
 \end{tabular}
 \caption{Effective observation time of the Crab Nebula subsamples in each of the NSB/hardware bins.}\label{tabTime}
 \end{table}
 
 \begin{figure}[t]
\includegraphics[width=\columnwidth]{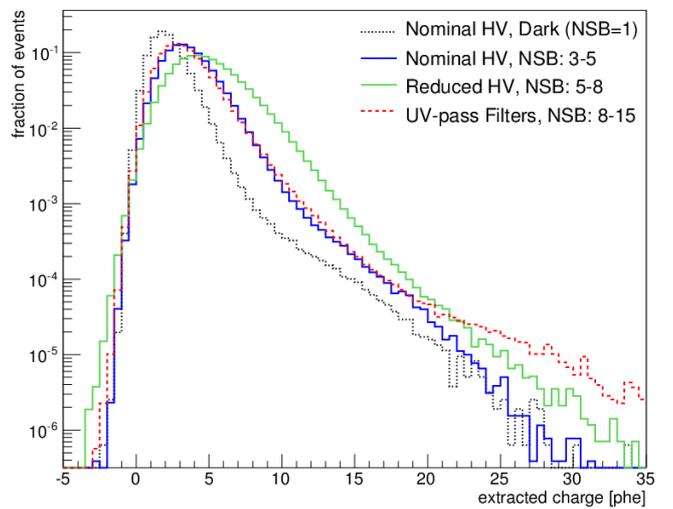}
\caption{Distributions of the pixel charge extracted with a sliding window for pedestal events (i.e., without signal) for different NSB/hardware conditions.}
\label{fig:pedchargedistr}
\end{figure}

\subsection{Analysis}\label{sec:analysis}

In this section we describe how moonlight affects the MAGIC data and how the analysis chain and MC simulations have been adapted.
The data have been analyzed using the standard MAGIC Analysis and Reconstruction Software (MARS, \cite{TrueMARS}) following the standard analysis chain described in \cite{upgrade2}, besides some modifications that were implemented to account for the different observation conditions.

\begin{table*}[t]
\centering
\begin{tabular}{| c | c | c | c | c |}
\hline
Sky Brightness & Hardware Settings & Pedestal Distr & Cleaning Level factors & Size Cut \\
               &             &  mean / rms    &  $\text{Lvl}_1$ / $\text{Lvl}_2$  &       \\
               
[$\textit{NSB}_{\text{Dark}}$]  &  &  [phe]  & [phe]     &   [phe]  \\
\hline
   \hline

     1 (Dark) & nominal HV & 2.0 / 1.0 &  6.0 / 3.5 &  50  \\
     1-2 & nominal HV & 2.5 / 1.2 & 6.0 / 3.5 &  60 \\
     2-3 & nominal HV & 3.0 / 1.3 & 7.0 / 4.5 &  80 \\
     3-5 & nominal HV & 3.6 / 1.5 & 8.0 / 5.0 & 110 \\
     5-8 & nominal HV & 4.2 / 1.7 & 9.0 / 5.5 & 150 \\
        \hline

     5-8 & reduced HV & 4.8 / 2.0 & 11.0 / 7.0 & 135 \\
     8-12 & reduced HV & 5.8 / 2.3 & 13.0 / 8.0 & 170 \\
     12-18 & reduced HV & 6.6 / 2.6 & 14.0 / 9.0 & 220 \\
        \hline

     8-15 & UV-pass filters & 3.7 / 1.6 & 8.0 / 5.0 & 100  \\
     15-30 & UV-pass filters & 4.3 / 1.8 & 9.0 / 5.5 & 135\\
   \hline
 \end{tabular}
 \caption{Noise levels of the Crab Nebula subsamples, adapted image cleaning levels and size cuts used for their analysis.}
 \label{tabNoise}
 \end{table*}

\subsubsection{Moonlight effect on calibrated data}\label{sec:calib}

After the trigger conditions are fulfilled, the signal of each pixel is recorded into a 30\,ns waveform. Then an algorithm looks over that waveform for the largest integrated charge in a sliding window of 3\,ns width, which is saved and later calibrated \citep{upgrade2}. In the absence of signal, the sliding window picks up the largest noise fluctuation of the waveform. The main sources of noise are the statistical fluctuations due to NSB photons, the PMT after pulses and the electronic noise. The noise due to background light fluctuations scales as the square root of the NSB (Poisson statistics). The after pulse rate is proportional to the PMT current, which increases linearly with the NSB. When the PMTs are operated under nominal HV, electronic noise has a similar level to the NSB fluctuation induced by a dark extragalactic FOV, which has no bright stars \citep{upgrade1}. For Crab dark observations, the brightness of the FOV ($\textit{NSB}_{\text{Dark}}$) is about 70\% higher than dark extragalactic FOV, and the NSB-related noise already dominates. Figure~\ref{fig:pedchargedistr} shows the distribution of extracted charge in photoelectrons (phe) for pedestal events (triggered randomly without signal) under four different observation conditions. During observations of the Crab Nebula under dark conditions the pedestal distribution has an RMS of $\sim $1\,phe and a mean bias of $\sim2$\,phe. The distribution is asymmetric with larger probability of upward fluctuation (induced by the sliding window method) and an extra tail at large signals ($>$8\,phe) produced by the PMT after pulses.

During moonlight observations, the noise induced by the NSB increases while the electronic noise remains constant (as long as the hardware settings remain unchanged). In fact, the electronic noise in terms of photoelectrons is proportional to the calibration constant, which depends on the hardware configuration of the observations. With reduced HV, all gains are lower, and hence the calibration constants increase resulting in higher electronic noise level in phe ($\sim$1.7) and, as a consequence, worse signal-to-noise ratio of integrated pulses. The transient time in PMTs also increases when the gain is lowered, but the delay in arrival time of pulses is $\sim$1 ns. The signal pulse is always well within the 30\,ns window and then the peak search method is not affected. During UV-pass filter observations PMTs are operated with nominal HV but some pixels are partially shadowed by the filter frame\footnote{The shadowing of the frame is important (blocking more than 40\% of the incoming light) for $\sim$7\% of the pixels.}. The camera flat-fielding, which makes all pixels respond similarly to the same sky light input, gives higher calibration constants to the shadowed pixels. Thus, electronic noise on those pixels is larger, while in contrast the NSB noise is strongly reduced by the filters. The relative contribution of the electronic to the total noise is then also higher during UV-pass filter observations. Table~\ref{tabNoise} shows the typical pedestal distribution mean and RMS for all the NSB/hardware bins.

The broader pedestal charge distribution has a double effect on the extraction of a real signal (Cherenkov light). If the signal is weak, the maximal waveform fluctuation may be larger than the Cherenkov pulse and the sliding window could select the wrong section. Then, the reconstructed pulse time is random and the signal is lost. If the signal is strong enough, the sliding window selects the correct region, the time and amplitude of the signal is just less precise (NSB does not induce a significant bias). Strong signals are almost not affected as their charge resolution is dominated by close to Poissonian fluctuations of the number of recorded phe.

\subsubsection{Moonlight-adapted image cleaning}

After the calibration of the acquired data, charge and timing information of each pixel is recorded. Most pixel signals contain only noise. The so-called sum-image cleaning \citep{upgrade2} is then performed to remove those pixels. In this procedure we search for groups of 4, 3 and 2 neighboring (4NN, 3NN, 2NN) pixels with a summed charge above a given level, within a given time window. The charge thresholds for 4NN-, 3NN-, 2NN-charge thresholds are set to $4 \times \text{Lvl}_1$, $3 \times 1.3 \times \text{Lvl}_1$, $2 \times 1.8 \times \text{Lvl}_1$, respectively, where $\text{Lvl}_1$ is a global factor adapted to the noise level of the observations. The time windows are kept fixed at 1.1\,ns, 0.7\,ns and 0.5\,ns, respectively, independent on the NSB level. Pixels belonging to those groups are identified as core pixels. Then all the pixels neighboring a core pixel that have a charge higher than a given threshold ($\text{Lvl}_2$) and an arrival time within 1.5\,ns with respect to that core pixel, are included in the image. In the MAGIC standard analysis \citep{upgrade2} the cleaning levels are set to $\text{Lvl}_1 = 6$\,phe and $\text{Lvl}_2=3.5$\,phe, which provide good image cleaning for any moonless-night observation. Higher cleaning levels would result in a higher energy threshold at the analysis level. In contrast, lower cleaning levels can also be used for dark extragalactic observation to push the analysis threshold as low as possible \citep{B0218}. The standard-analysis cleaning levels are then a compromise between robustness and performance, optimized to be used for any FOV, galactic or extragalactic, under dark and dim moonlight conditions.

During moonlight observations the background fluctuations are higher and the cleaning levels must be increased accordingly. Those levels were modified to ensure that the fraction of pedestal events that contain only noise and survive the image cleaning is lower than 10\%. They were optimized for every NSB/hardware bin independently to get the lowest possible analysis threshold for every bin. The optimized cleaning levels for each bin are shown in Table~\ref{tabNoise}. The time window widths were not modified for reduced HV observations, because the variations in the PMTs response are expected to be very small.

We do not use variable cleaning levels that would automatically scale as a function of the noise because the MAGIC data reconstruction is based on comparison with MC simulations, which must have exactly the same cleaning levels as the data. During moonlight observations, the noise level is continuously changing, so it is not realistic to fine tune our MC for every observation. Instead we create a set of MC simulations for every NSB/hardware bin with fixed noise and cleaning levels.

\subsubsection{Moon-adapted Monte Carlo simulations}\label{sec:MC}

\begin{figure*}[t]
\centering
\includegraphics[width=1.5\columnwidth]{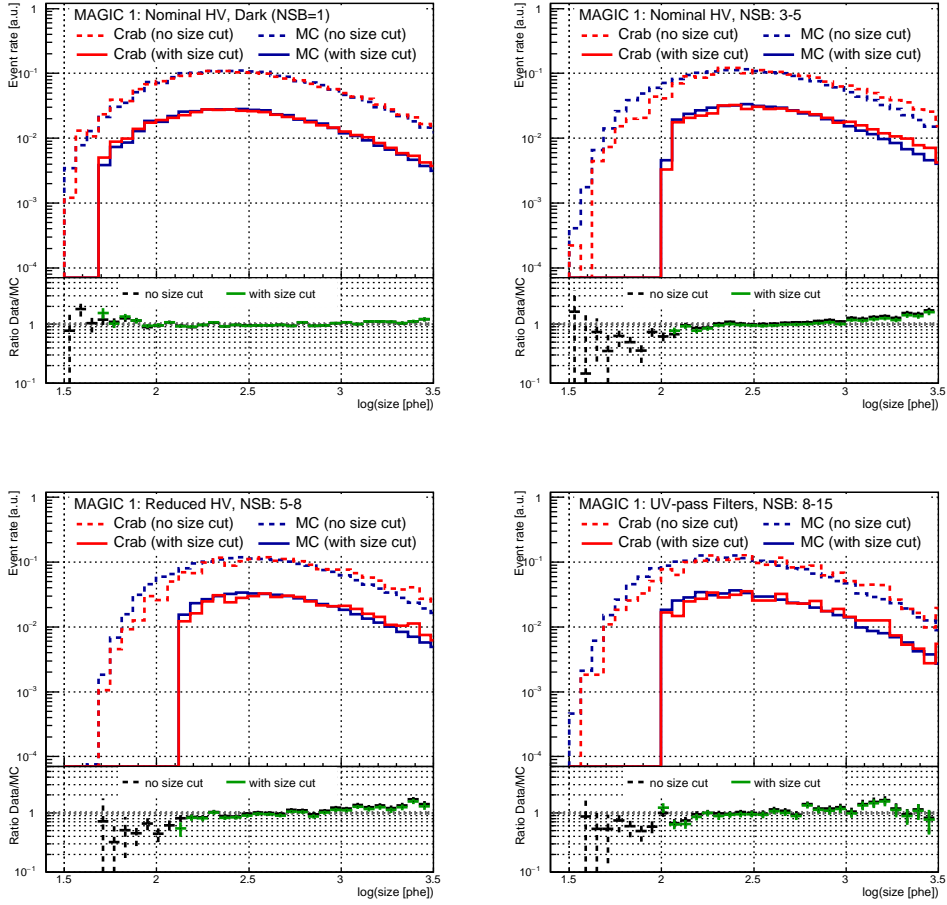}
\caption{Comparison between MAGIC 1 data (red) and MC $\gamma$-ray (blue) image size distributions for different NSB/hardware bins. Data distributions are composed by excess events within a 0.14$^\circ$ circle around the Crab Nebula position. MC distributions were simulated with the same energy distribution as the Crab Nebula spectrum reported in \cite{upgrade2}. In dashed and solid lines the distributions before and after applying the optimized size cuts are shown. Distributions with and without size cuts were normalized to different values for a better visualization. Lower panels show the ratio of the data distributions to the MC ones. }\label{fig:SizeM1}
\end{figure*}


MC simulations have mainly two functions in the MAGIC data analysis chain. A first sample (train sample) is used to build look-up tables and multivariate decision trees (random forest), which are employed for the energy and direction reconstruction and gamma/hadron separation \citep{upgrade2}. A second, independent sample (test sample) is used for the telescope response estimation during the source flux/spectrum reconstruction.

We prepared MC samples adapted for every NSB/hardware bin. For nominal and reduced HV settings, we used the standard MAGIC MC simulation chain with additional noise to mimic the effect of moonlight (and reduced HV). The noise is injected after the calibration at the pixel signal level. First we model the noise distribution in a given integration window of 3\,ns that would produce the same pedestal charge distribution than the one obtained during observations (see Figure \ref{fig:pedchargedistr}) using the sliding window search method described in Section \ref{sec:calib}. We then extract a random value from the modeled noise distribution and add it to the extracted signal of the MC event. If the modified signal is larger than a random number following the pedestal charge distribution, this new value becomes the new charge and a random jitter is added to the arrival time (depending on the new signal/noise ratio). If the random pedestal signal is larger it means that the sliding window caught a spurious bump larger than the signal itself, then the pixel charge is set to this fake signal and the arrival time is chosen randomly according to the pedestal time distribution. This method allows us to adapt our MC to any given NSB without reprocessing the full telescope simulation and data calibration. In the case of the UV-pass filter observations, additional modifications on the simulation chain were implemented to include the filter transmission and the shadowing produced by the frame ribs.

We did not simulate the effect of the moonlight on the trigger because it is very difficult to reproduce the behavior of the IPRC, which control the pixel DTs (see section~\ref{sec:trigger}). Instead, simulations were performed using the standard dark DTs and we later applied cuts on the sum of charge of pixels surviving the image cleaning (image size) on each telescope. This size cut acts as a software threshold and it is optimized bin-wise as the minimal size for which the data and MC distributions are matching. Even in the absence of moonlight a minimum cut in the total charge of the images is applied, as potential $\gamma$-ray events with lower sizes are either harder to reconstruct or to distinguish from hadron-induced showers \citep{upgrade2}.  The used size cuts are given Table~\ref{tabNoise}. Figure \ref{fig:SizeM1} compares size distributions of MC $\gamma$-ray events (simulated with the spectrum of the Crab Nebula reported in \citealt{upgrade2}) with those of the observed excess events within a 0.14$^\circ$ circle from the Crab Nebula.

\section{Performance}

In this section we evaluate how moonlight and the use of different hardware configurations affect the main performance parameters of the MAGIC telescopes.

\subsection{Energy threshold}

\begin{figure}[t]
\includegraphics[width=\columnwidth]{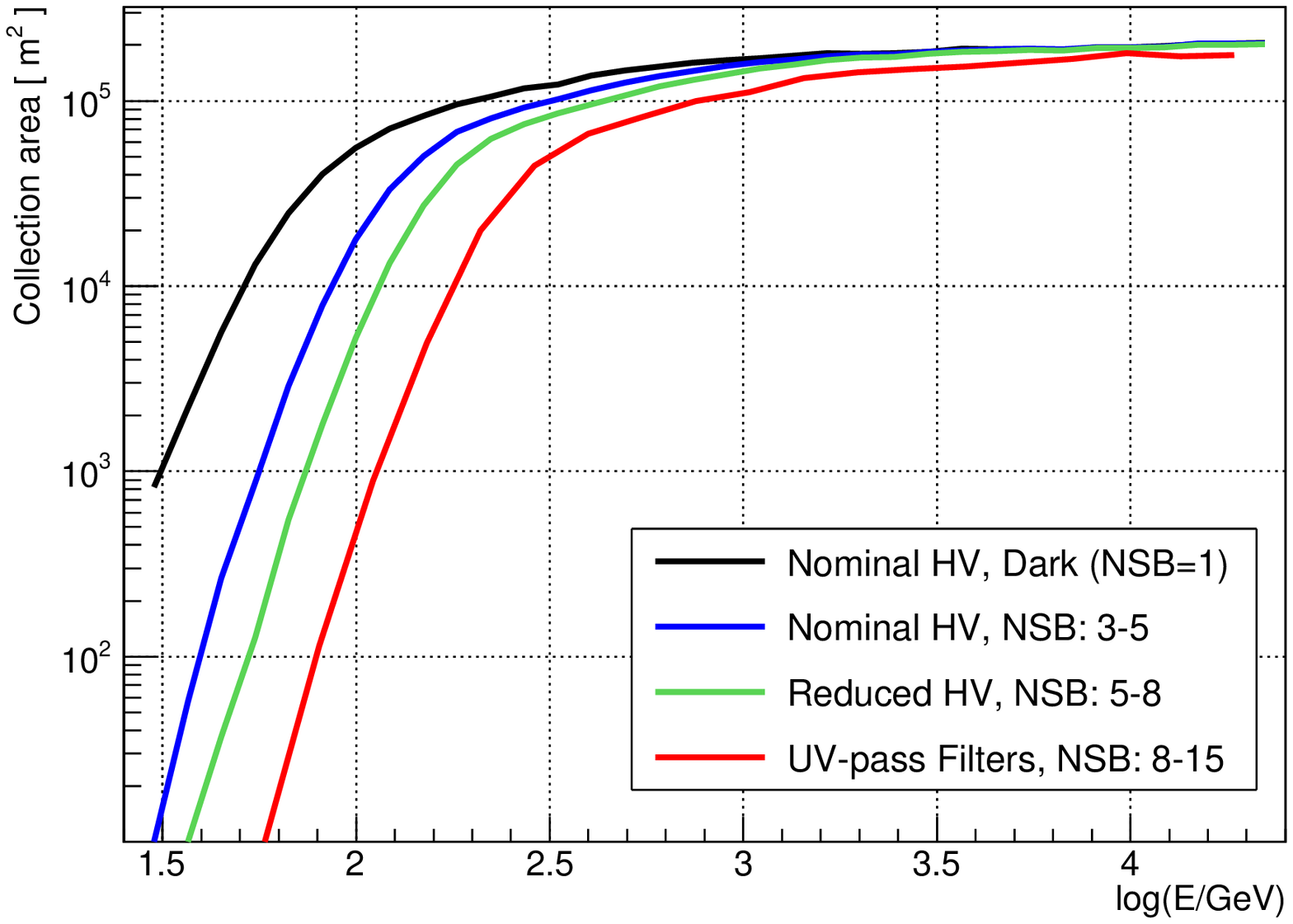}
\caption{Effective collection area at reconstruction level for zenith angles below 30$^\circ$ for four different observation conditions: Dark conditions with nominal HV (black),  3-5~$\times  \textit{NSB}_{\text{Dark}}$ with nominal HV (blue), 5-8~$\times  \, \textit{NSB}_{\text{Dark}}$ with reduced HV (green) and 8-15~$\times  \textit{NSB}_{\text{Dark}}$ with UV-pass filters (red). The optimized cleaning levels and size cuts from Table~\ref{tabNoise} were used to produce these plots.}\label{fig:CollAreaExample}
\end{figure}

The energy threshold of IACT telescopes is commonly defined as the peak of the differential event rate distribution as a function of energy. It is estimated from the effective collection area as a function of the energy, obtained from $\gamma$-ray MC simulations, multiplied by the expected $\gamma$-ray spectrum, which is typically (and also in this work) assumed to be a power-law with a spectral index of $-2.6$. It can be evaluated at different stages of the analysis. The lowest threshold corresponds to the trigger level, which reaches $\sim50$ GeV during MAGIC  observations in moonless nights at zenith angles below 30$^\circ$ \citep{upgrade2}. It naturally increases during moonlight observations, as the DTs are automatically raised by the IPRC (see Section \ref{sec:trigger}). As explained in section~\ref{sec:analysis}, our MC simulations do not reproduce the complex behavior of the trigger during such observations. Here we evaluate then the energy threshold at a later stage, after image cleaning, event reconstruction and size cuts (reconstruction level), for which a good matching between real data and MC is achieved.


The effective collection area at the reconstruction level as a function of the energy for four different NSB/hardware situations are shown in Figure~\ref{fig:CollAreaExample}. In all four curves two regimes can be identified: one, at low energies, which is rapidly increasing with the energy and another, towards high energies, which is close to a plateau. As expected, the dark-sample analysis presents the largest effective area along the full energy range. The degradation due to moonlight is more important at the lowest energies, where the Cherenkov images are small and dim. The higher the size cuts and cleaning levels, the higher the energy at which the plateau is achieved. In the case of UV-pass filter observations, the used cleaning levels and size cuts are lower (in units of phe) than the ones applied during reduced HV data analysis, but due to the filter transmission, the plateau is reached at even higher energies. Above $\sim$1\,TeV the effective area is almost flat for the four studied samples and the effect of Moon analysis is very small (below $\sim$10\%).

\begin{figure}[t]
\includegraphics[width=\columnwidth]{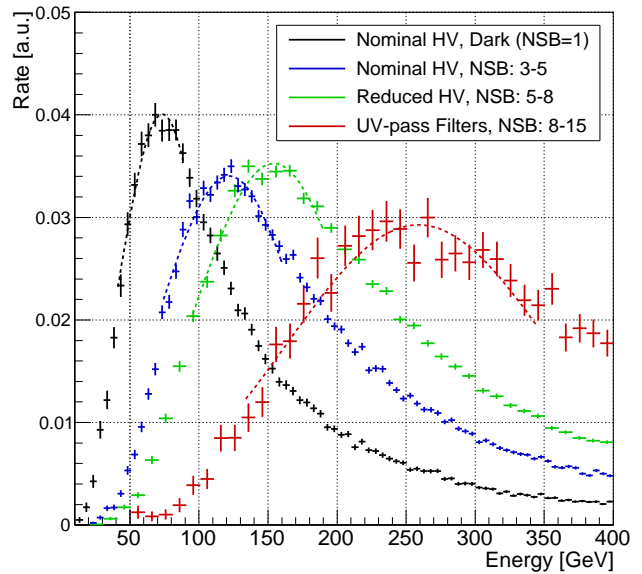}
\caption{Rate of MC $\gamma$-ray events that survived the image cleaning and a given quality size cut for an hypothetical source with an spectral index of $-2.6$ observed at zenith angles below 30$^\circ$. The four curves correspond to different observation conditions: Dark conditions with nominal HV (black),  3-5~$\times  \textit{NSB}_{\text{Dark}}$ with nominal HV (blue), 5-8~$\times  \, \textit{NSB}_{\text{Dark}}$ with reduced HV (green) and 8-15~$\times  \textit{NSB}_{\text{Dark}}$ with UV-pass filters (red). Dashed lines show the gaussian fit applied to calculate the energy threshold on each sample.}\label{fig:EthExample}
\end{figure}

\begin{figure}[t]
\includegraphics[width=\columnwidth]{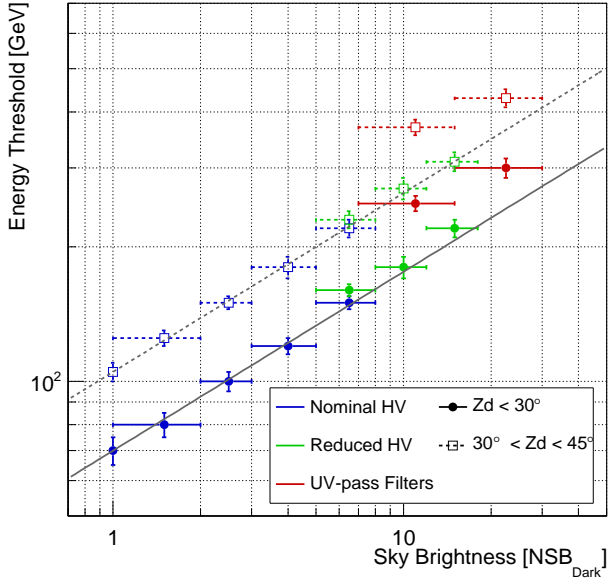}
\caption{Energy threshold at the event reconstruction level as a function of the sky brightness for observations with nominal HV (black), reduced HV (green) and UV-pass filters (red) at zenith angles below 30$^\circ$ (filled circles, solid lines) and between 30$^\circ$ and 45$^\circ$ (empty squares, dashed lines). Gray lines represent the approximation given by equation \ref{eq:ThrFit} for zenith angles below 30$^\circ$ (solid) and between 30$^\circ$ and 45$^\circ$ (dashed).}\label{fig:EthNSB}
\end{figure}

The degradation of the effective area at low energies is directly translated into an increase of the energy threshold, as can be seen in Figure \ref{fig:EthExample}, where the differential rate plots for the same four NSB/hardware cases are shown. The energy threshold at reconstruction level is estimated by fitting a Gaussian distribution in a narrow range around the peak of these distributions\footnote{Note that in those distributions the peak is broad, which means that it is possible to obtain scientific results with the telescopes below the defined threshold.}. In Figure \ref{fig:EthNSB} we show the obtained energy threshold as a function of the sky brightness for different hardware configurations at low ($<30^\circ$) and medium ($30^\circ - 45^\circ$) zenith angles\footnote{Here we compute an average over a relatively wide zenith range, but energy threshold dependence with the zenith angle is stronger for medium zenith angles (see Figure 6 in \cite{upgrade2})}. For low zenith angles it goes from $\sim$70\,GeV in the absence of moonlight to $\sim$300\,GeV in the brightest scenario considered. For medium zenith angles, the degradation is similar from $\sim$110\,GeV to  $\sim$500\,GeV. The degradation of the energy threshold $E_{\text{th}}$ as a function of the NSB level can be roughly approximated, for nominal HV and reduced HV data, by

\begin{equation}\label{eq:ThrFit}
E_{\text{th}}(\textit{NSB}) = E^{\text{Dark}}_{\text{th}} \times \left(\dfrac{\textit{NSB}}{\textit{NSB}_{\text{Dark}}}\right)^{0.4}
\end{equation}

Where $E^{\text{Dark}}_{\text{th}}$ is the energy threshold during dark Crab Nebula observations.
At the same NSB level, reduced HV data have a slightly higher energy threshold than nominal HV data due to higher electronic noise in phe units, while the UV-pass-filter energy threshold is significantly higher ($\sim$40\%) than the one of reduced HV data without filters. The energy threshold increase with filters is due to the lower photon statistic (the same shower produces less phe). This degradation is reduced at higher NSBs (i.e. higher energies), where larger image sizes make the photon statistic less important than the signal-to-noise ratio in the energy threshold determination.

\subsection{Reconstruction of the Crab Nebula spectrum}\label{sec:CrabSpectra}

\subsubsection{Standard cleaning}

MAGIC data are automatically calibrated with the standard analysis chain optimized for dark observations. Most of the analyses start from high level data, after image cleaning and event reconstruction. When dealing with moonlight data an adapted analysis is in principle required, as described in Section \ref{sec:analysis}. However, the effect of weak moonlight can be almost negligible and the data can be processed following the standard chain. Here we want to determine which is the highest NSB level for which the standard analysis provides consistent results, within reasonable systematic uncertainties, with respect to those obtained with the dark reference sample.

\begin{figure}[h]
\centering
\includegraphics[trim=0cm 0cm 1.5cm 0cm, clip=true, width=\columnwidth]{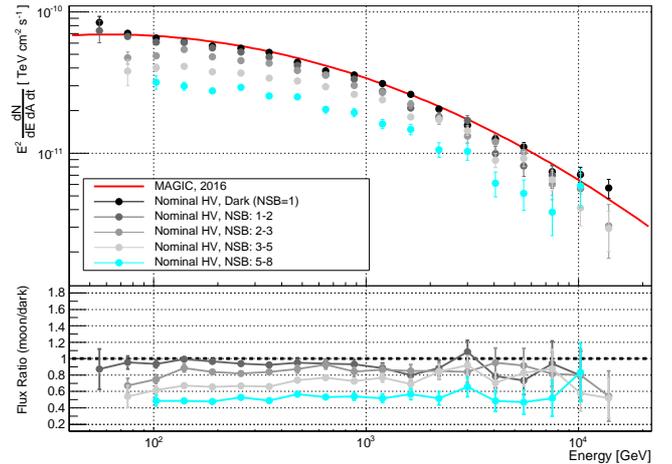}
\caption{Spectral energy distribution of the Crab Nebula obtained for different NSB levels (given in units of $\textit{NSB}_{\text{Dark}}$)  using the standard analysis, compared to the result obtained previously by MAGIC (best log-parabola fit in red solid line, \citep{upgrade2}). The lower panel shows the ratio of the fluxes measured under moonlight to the ones measured in dark conditions.}\label{fig:CrabSEDStd}
\end{figure}

To answer this question we attempted to reproduce the Crab Nebula spectrum by applying the standard analysis, including standard dark MC for the train and test samples, to our moonlight data taken with nominal HV. To minimize systematic uncertainties we use typical selection cuts with 90\% $\gamma$-ray efficiency for the $\gamma$-ray/hadron separation and sky signal region radius \citep{upgrade2}. The obtained Crab Nebula spectral energy distributions (SEDs) are shown in figure~\ref{fig:CrabSEDStd} for 1-8~$\times  \, \textit{NSB}_{\text{Dark}}$. The image size cuts described in Section \ref{sec:MC} were applied to produce these spectra. The SED obtained using data with 1-2~$\times  \, \textit{NSB}_{\text{Dark}}$ is compatible, within errors, with the one obtained with dark data. This shows that the standard analysis is perfectly suitable for this illumination level. For brighter NSB conditions the reconstructed spectra are underestimated. With 2-3~$\times  \, \textit{NSB}_{\text{Dark}}$, the data-point errors above $\sim$130\,GeV are below $\sim$20\% while with 5-8~$\times  \, \textit{NSB}_{\text{Dark}}$ the reconstructed flux falls below $\sim$50\% at all energies. Thus, the standard analysis chain can be still used for weak moonlight at the price of additional systematic bias (10\% for 1-2~$\times  \, \textit{NSB}_{\text{Dark}}$ and 20\% for 2-3~$\times  \, \textit{NSB}_{\text{Dark}}$) but for higher NSB levels a dedicated Moon analysis is mandatory.

\subsubsection{Custom analysis}

\begin{figure}
\includegraphics[trim=0cm 0cm 1.5cm 0cm, clip=true, width=\columnwidth]{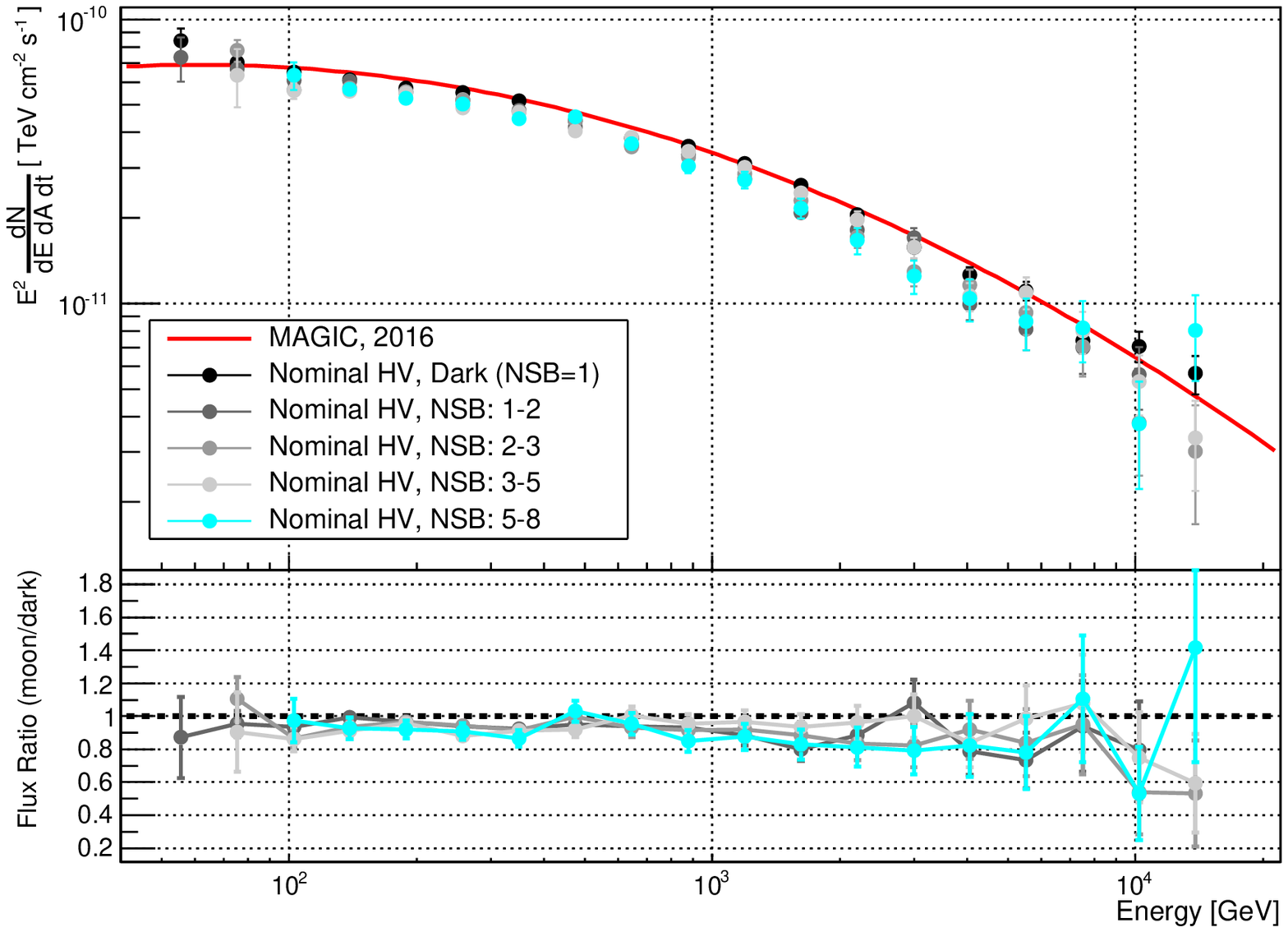}
\includegraphics[trim=0cm 0cm 1.5cm 0cm, clip=true, width=\columnwidth]{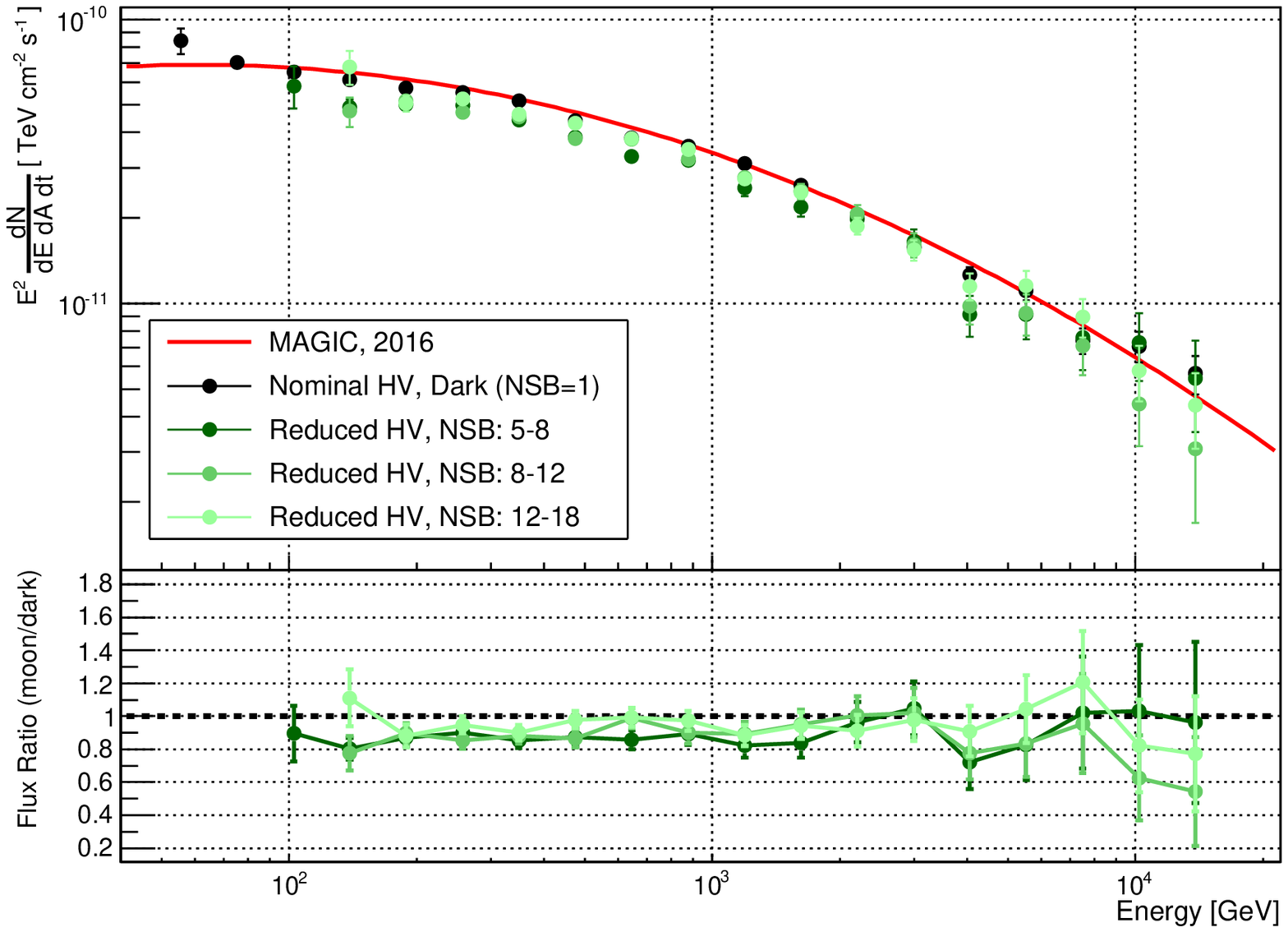}
\includegraphics[trim=0cm 0cm 1.5cm 0cm, clip=true, width=\columnwidth]{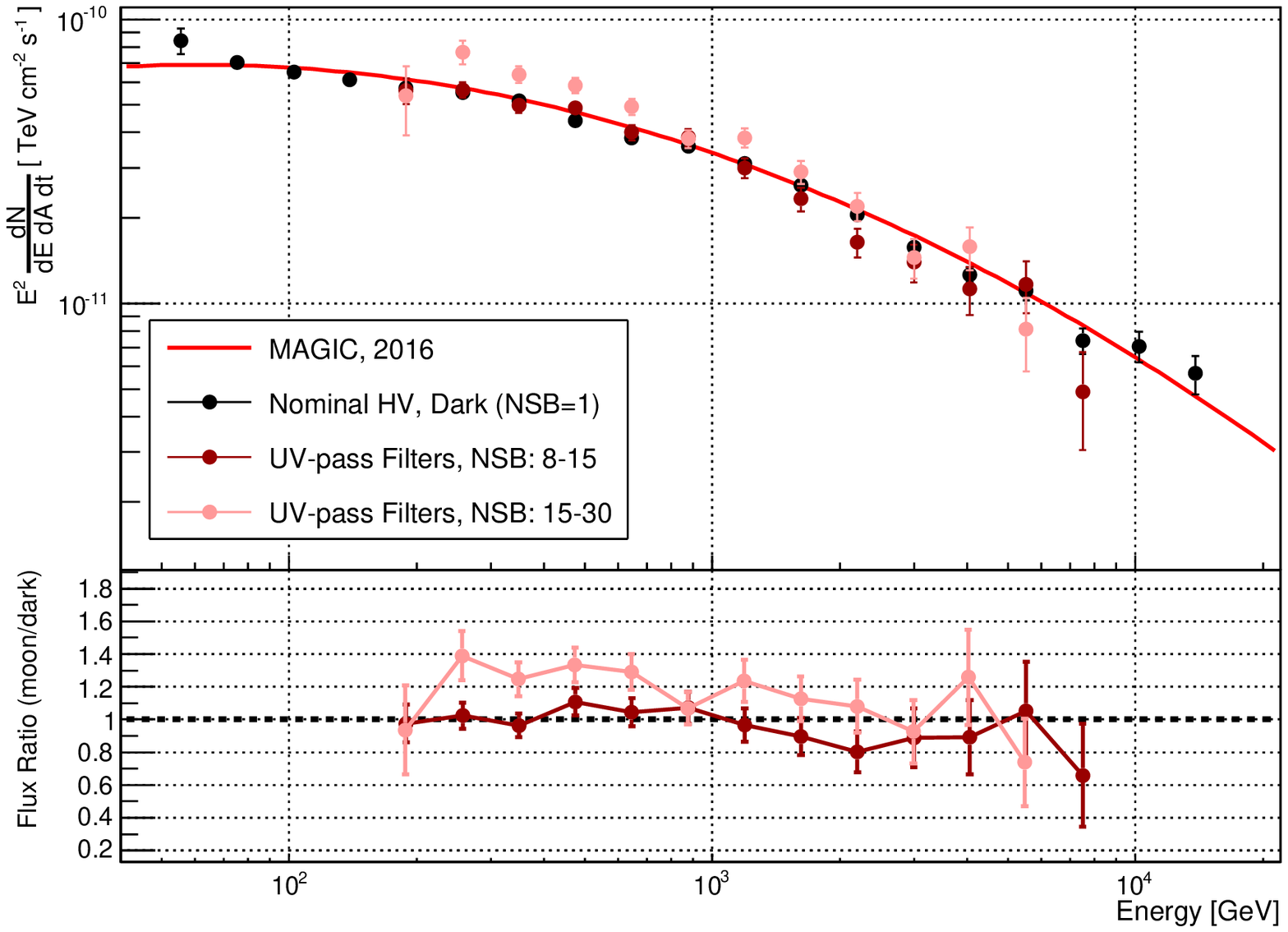}
\caption{Spectral energy distribution of the Crab Nebula obtained for different NSB levels (given in units of $\textit{NSB}_{\text{Dark}}$, coloured dots) using the dedicated Moon analysis for nominal HV (top), reduced HV (centre) and UV-pass filters (bottom) data. For comparison the result obtained with the dark sample using standard analysis in this work (black dots) and previously published by MAGIC (red solid line, \citep{upgrade2}) are shown in every panel. The bottom sub-panels show the ratio of the fluxes measured under moonlight to the flux measured under dark conditions.}\label{fig:CrabSEDMoon}
\end{figure}

Figure~\ref{fig:CrabSEDMoon} shows the spectra of the Crab Nebula obtained after applying the dedicated Moon analysis (dedicated MC, cleaning levels and size cuts) described in Section \ref{sec:analysis} to each data set. In almost all the cases the fluxes obtained are consistent within $\pm$20\% with the one obtained under dark conditions, at least up to 4\,TeV. The only exception is the brightest NSB bin (UV-pass filters data up to 30 $\times \textit{NSB}_{\text{Dark}}$) where the ratio of the flux to the dark flux gets slightly above $\sim$30\% at energies between about 400 and 800\,GeV. It is also interesting to notice how the spectrum reconstruction improves when the dedicated moon analysis is performed by comparing the spectra obtained for the nominal HV samples in Figures \ref{fig:CrabSEDStd} and \ref{fig:CrabSEDMoon}.

\subsection{Angular resolution}\label{sec:AngRes}

\begin{figure*}[t]
\centering
\includegraphics[width=\columnwidth]{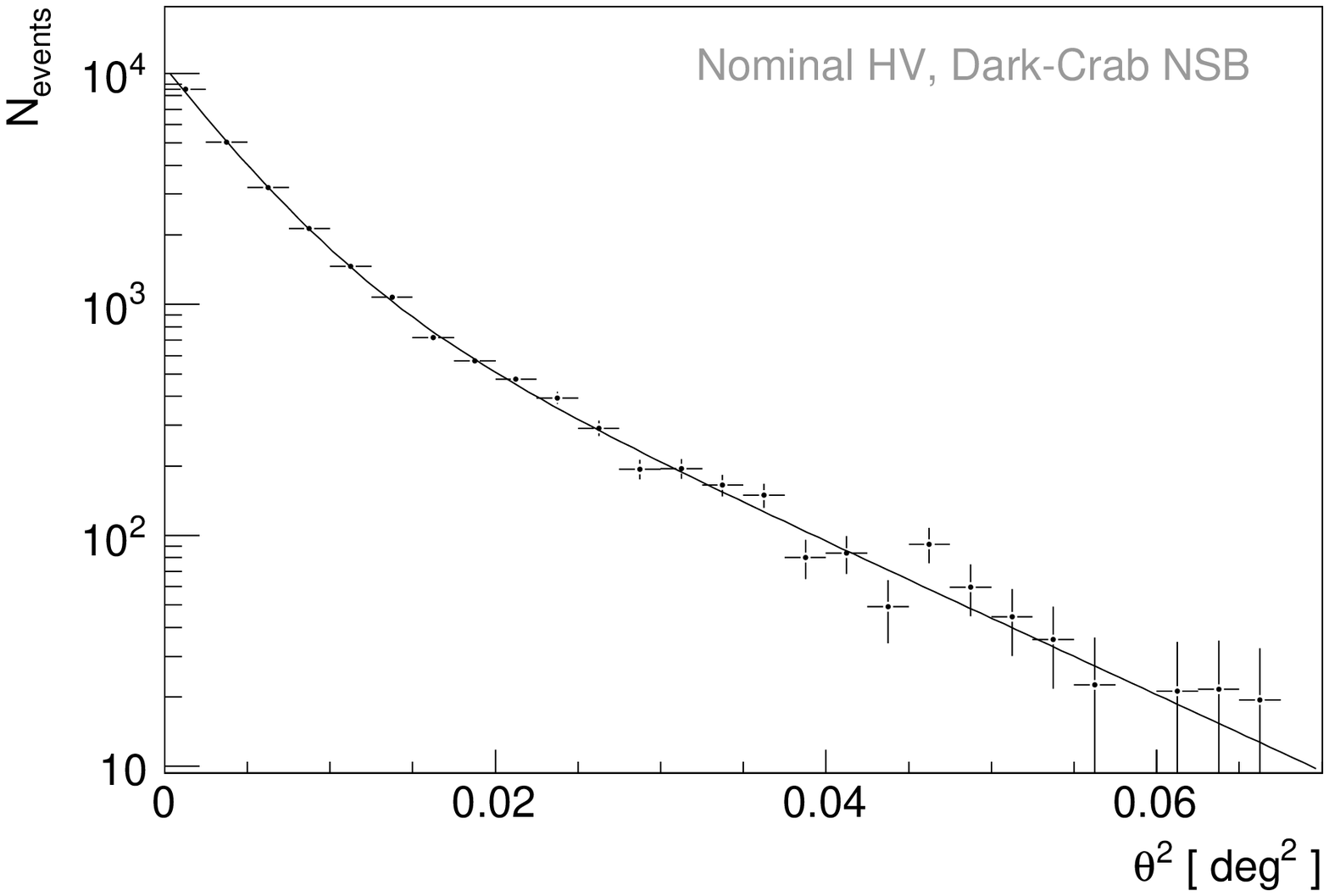}
\includegraphics[width=\columnwidth]{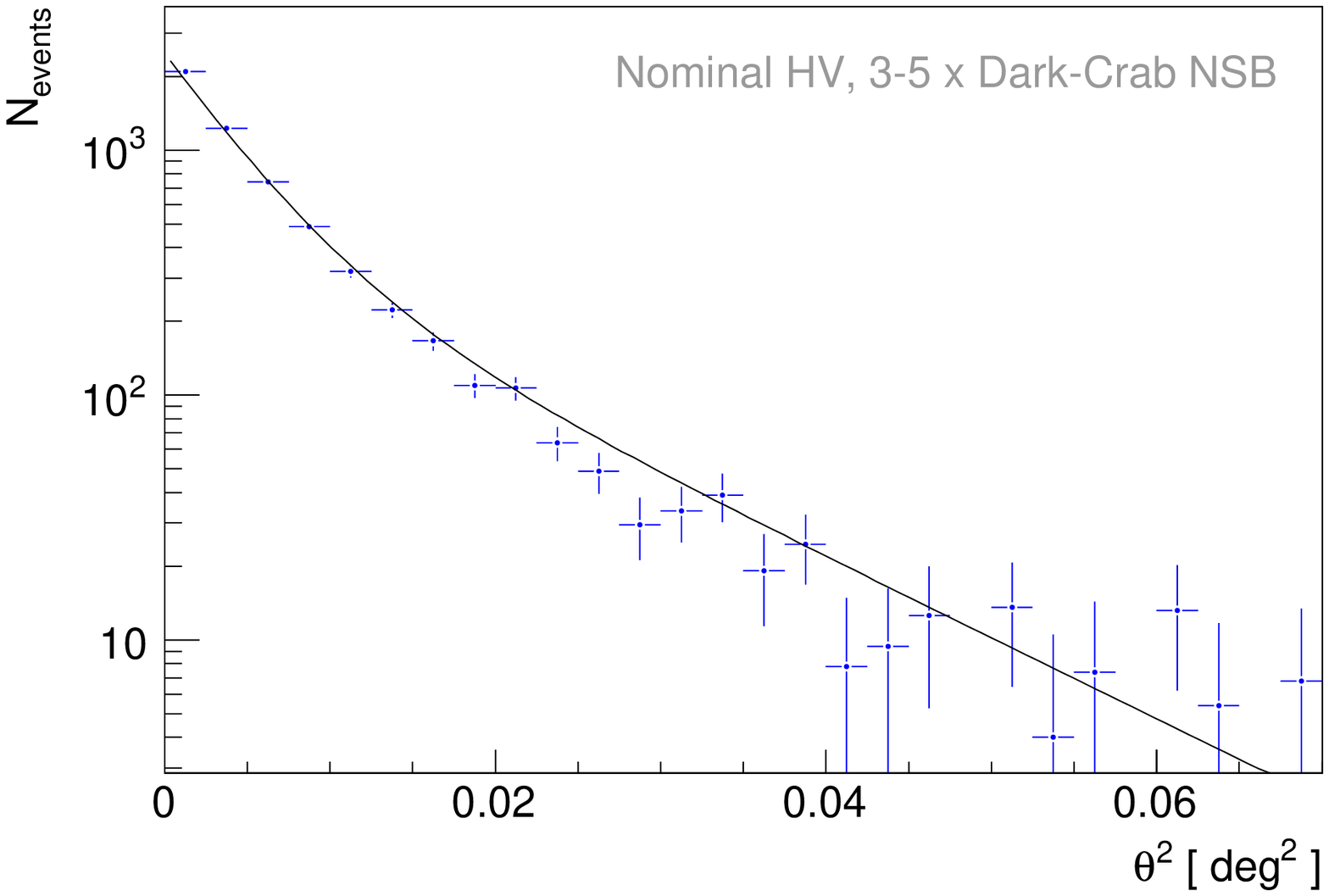}
\includegraphics[width=\columnwidth]{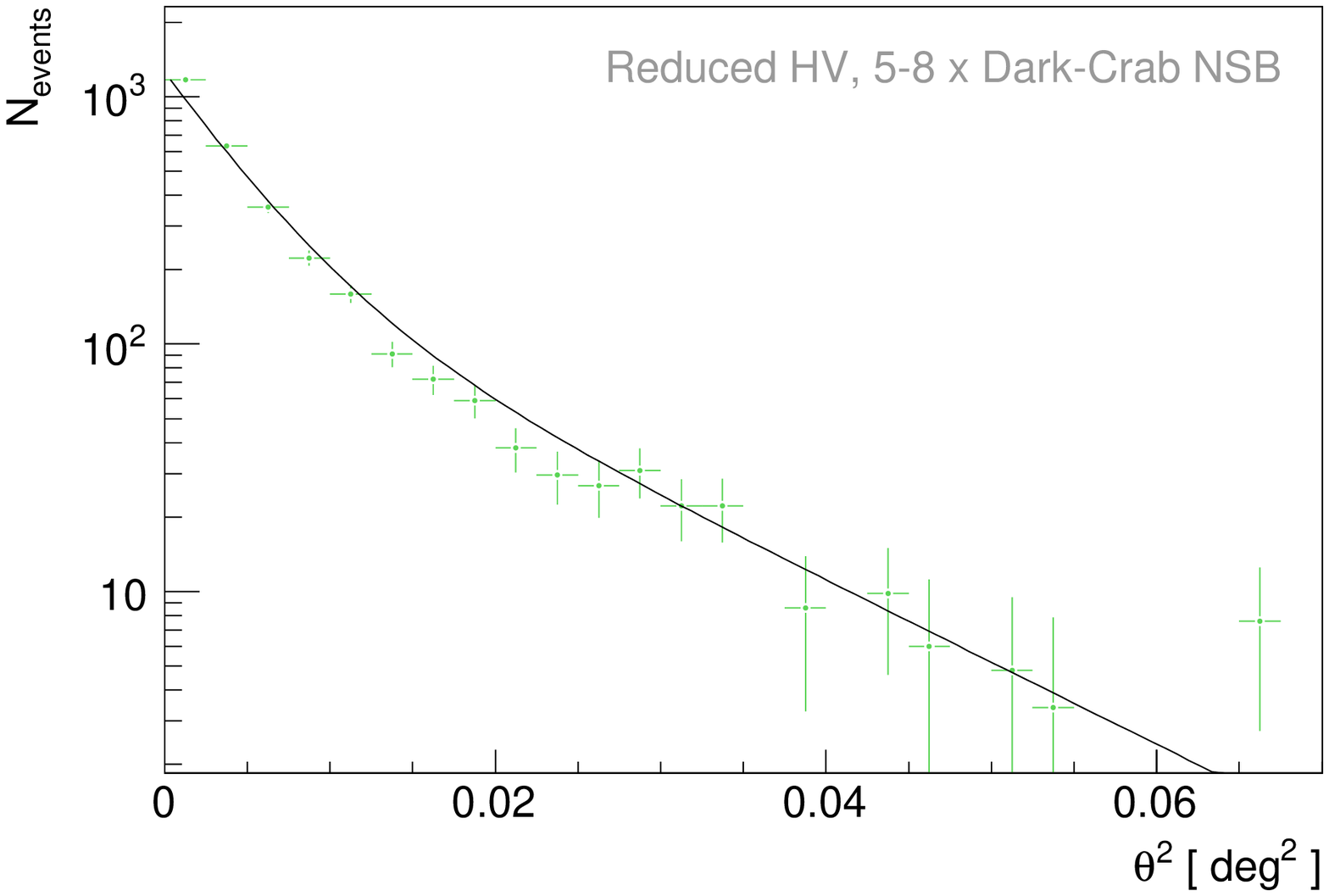}
\includegraphics[width=\columnwidth]{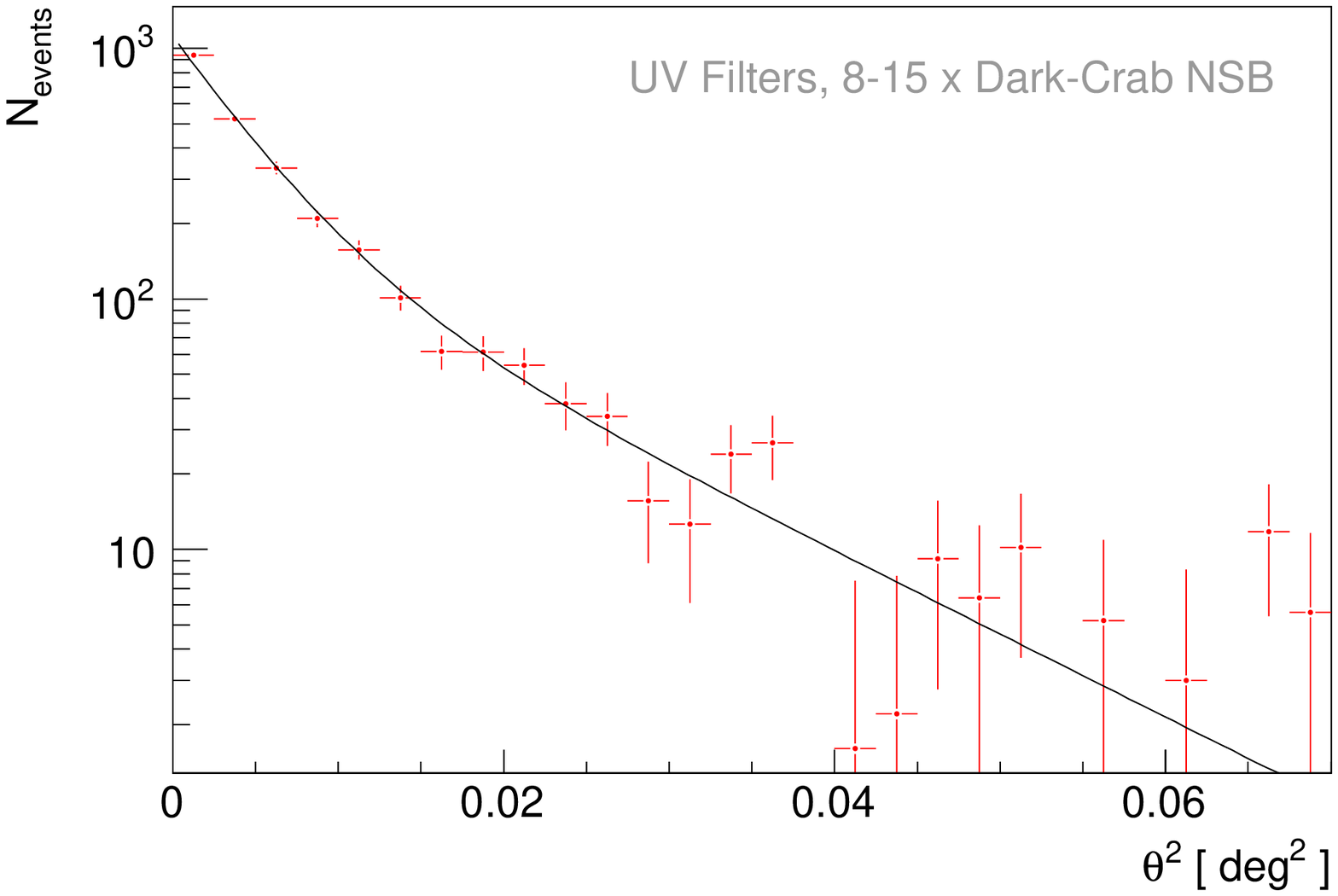}
\caption{ $\theta^2$ distribution of excess events ($\gamma$-ray events) with an estimated energy above 300\,GeV for the usual four cases studied: Dark (NSB~=~1), nominal HV NSB: 3-5, reduced HV NSB: 5-8, UV-pass filters NSB: 8-15 (NSB in $\textit{NSB}_{\text{Dark}}$ units). The solid black lines show the PSF fit (double-exponential) obtained with the dark sample.}\label{fig:angRes}
\end{figure*}

The reconstruction of the $\gamma$-ray arrival direction could be affected in two ways by moonlight. Firstly, as already discussed, it induces more background noise that affects the quality of the recorded images. Secondly the moonlight can disturb the tracking monitor of the telescope, which is based on a star-guiding system \citep{Riegel2005ICRC}. An eventual mispointing is ruled out by checking that for every NSB/hardware bin the center of the 2D-skymap event excess distribution (obtained with a Gaussian fit) is well within a 0.02$^\circ$ circle around the actual Crab Nebula position as expected from the pointing accuracy of MAGIC \citep{upgrade2}. To study the possible degradation of the point spread function (PSF), we compare the $\theta^2$ distribution obtained for Crab data taken under moonlight and under dark conditions, $\theta$ being the angular distance between the Crab Nebula position and the reconstructed event arrival direction. As explained in \citep{upgrade2}, this distribution can be well fitted by a double exponential function. Figure~\ref{fig:angRes} shows the $\theta^2$ distribution of events with estimated energy above 300\,GeV and $\gamma$-ray/hadron separation cut corresponding to 90\% $\gamma$-ray efficiency for four representative NSB/hardware bins. For all the NSB/hardware bins the $\theta^2$ distribution above the energy threshold is in good agreement with the PSF obtained under dark conditions. The angular resolution does not seem to be significantly affected by moonlight.

\subsection{Sensitivity}
As shown in previous sub-sections, moonlight observations are perfectly apt for bright $\gamma$-ray sources such as the Crab Nebula, whose spectrum and direction can be well reconstructed, with the only drawback being a higher energy threshold with respect to the one obtained in dark observations. However, one may wonder how the performance for the detection of weak sources is affected by moonlight, which may degrade the $\gamma$-ray/hadron separation power. To study this potential effect, we computed the minimal $\gamma$-ray flux that MAGIC can detect in 50\,h of observation, from $\gamma$-ray and background event rates obtained with the Crab Nebula samples analyzed in this work, following the method described in \citep{upgrade2} \footnote{The sensitivity is defined as the integral flux above an energy threshold giving $N_{\text{excess}} / \sqrt{N_{\text{bgd}}} = 5$, where $N_{\text{excess}}$ is the number of excess events and $N_{\text{bgd}}$ the number of background events, with additional constraints: $N_{\text{excess}} > 10$ and $N_{\text{excess}} > 0.05 N_{bgd}$.}.
For each NSB/hardware bin, the $\gamma$-ray and background rates are obtained for several analyses achieving different energy thresholds. 
Each analysis corresponds to a set of cuts in the image size and reconstructed energy as well as previously optimized $\gamma$-ray/hadron separation cuts. The analysis-level energy threshold is estimated by applying the same set of cuts to a $\gamma$-ray MC sample simulated with the same energy spectrum as the Crab Nebula and re-weighted to reproduce the same zenith-angle distribution as for the observations. 

\begin{figure}
\includegraphics[width=\columnwidth]{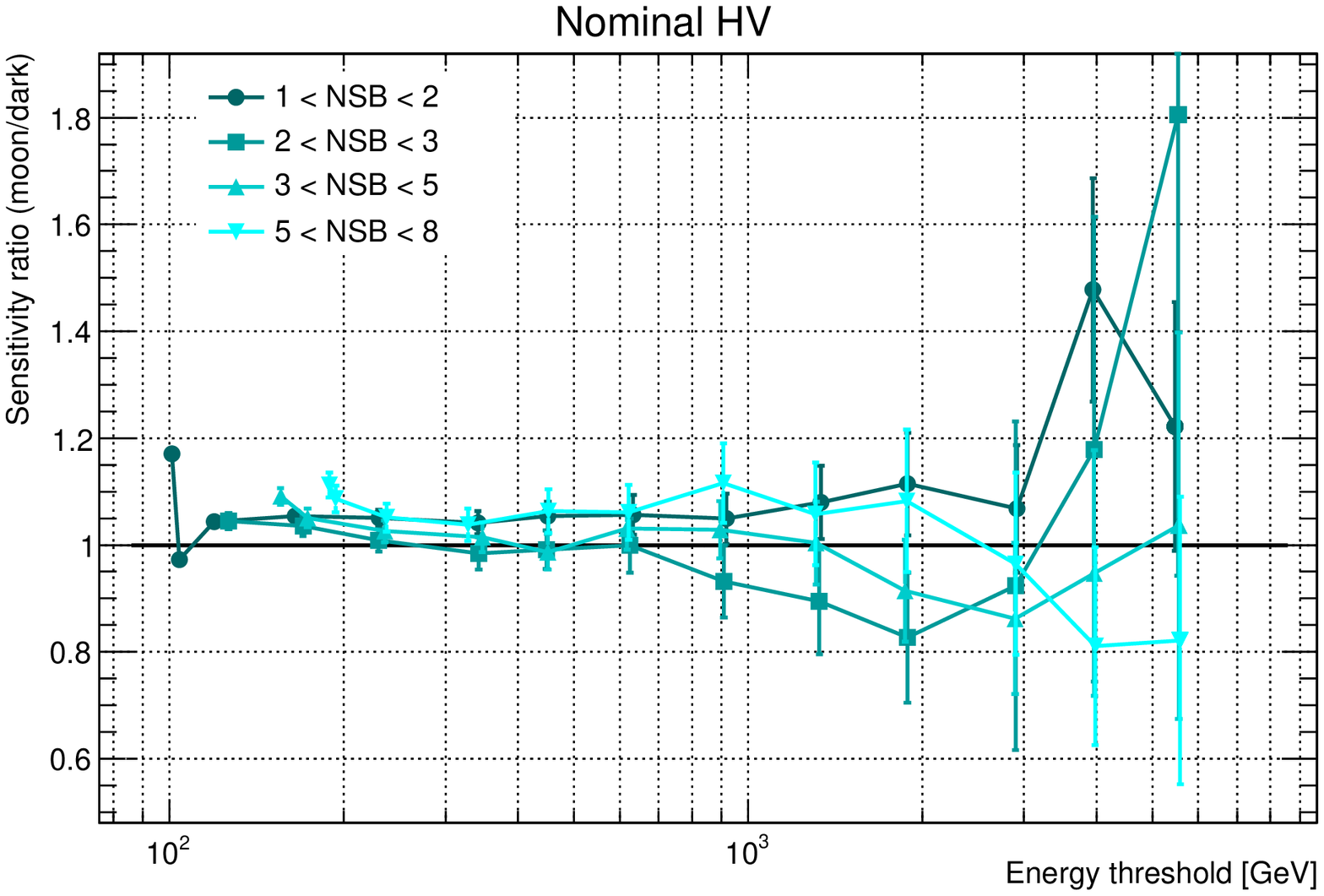}
\includegraphics[width=\columnwidth]{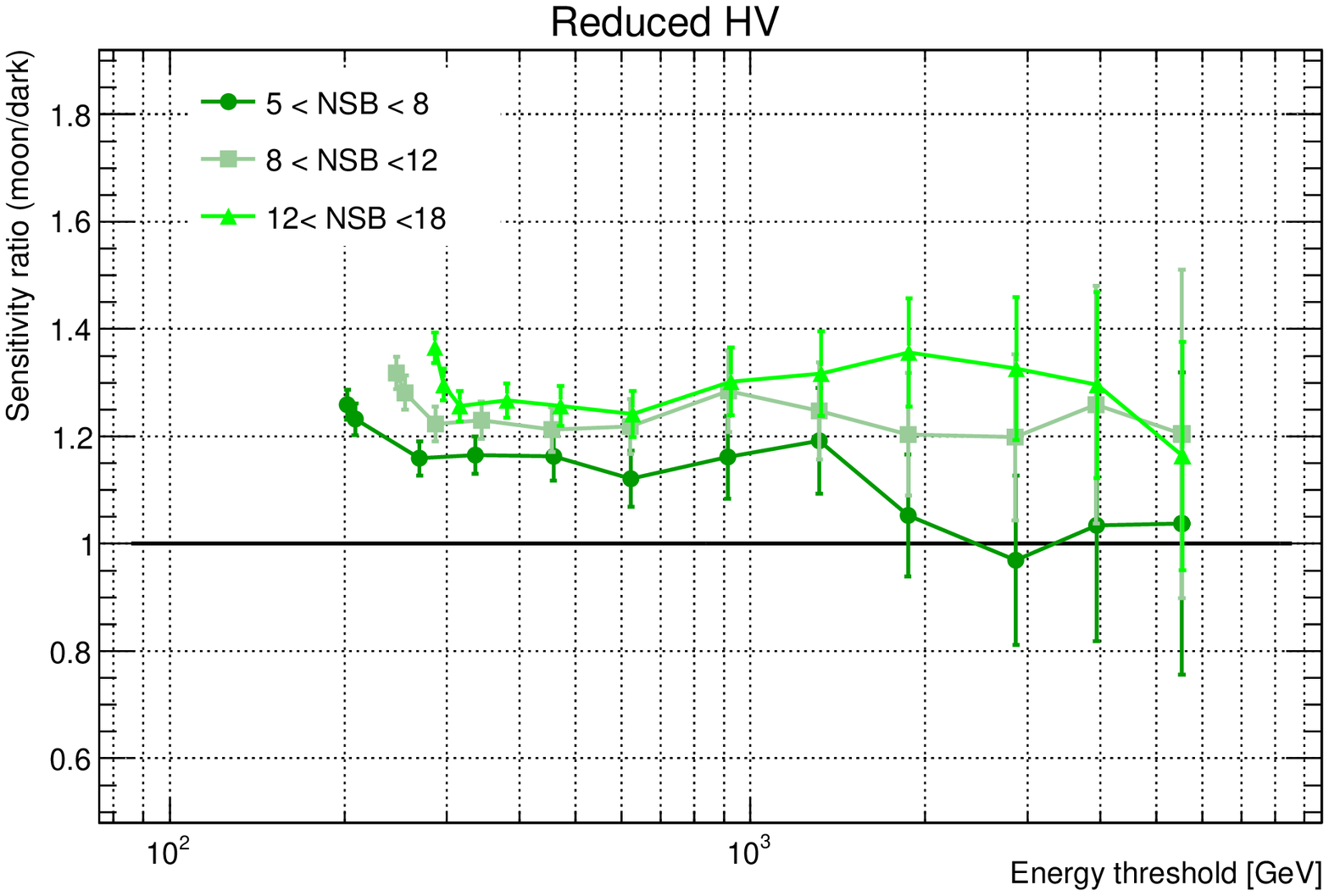}
\includegraphics[width=\columnwidth]{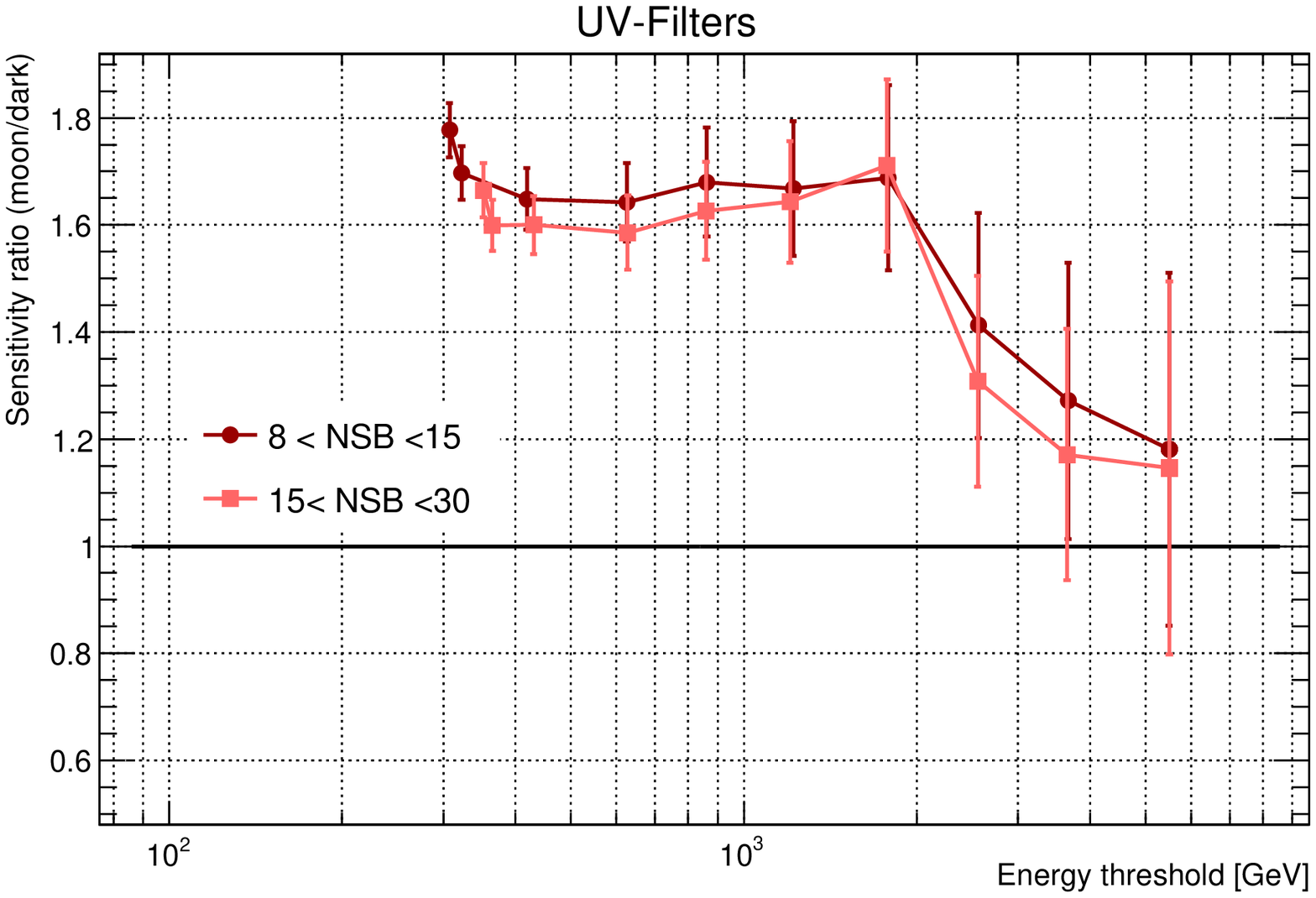}
\caption{Ratio of the integral sensitivity under moonlight to the dark sensitivity as a function of the analysis energy threshold, for nominal HV (top), reduced HV (middle) and UV-pass filter (bottom) data. The NSB levels are given in unit of $\textit{NSB}_{\text{Dark}}$}\label{fig:Sens}
\end{figure}

To accumulate enough data in every NSB/hardware bin, we use data from a large zenith angle range going from 5$^\circ$ to 45$^\circ$. As the sensitivity and energy threshold depend strongly on the zenith angle and data sub-samples have different zenith angle distributions, the performances are corrected to correspond to the same reference zenith-angle distribution (average of all the data). To visualize the degradation caused by moonlight, the integral sensitivity computed for each NSB/hardware bin is divided by the one obtained under dark conditions at the same analysis-level energy threshold. The obtained sensitivity ratios are shown in Figure \ref{fig:Sens} as a function of the energy threshold. The Moon data taken with nominal HV provide a sensitivity only slightly worse than the one obtained using dark data. The sensitivity degradation is constrained to be less than 10\% below 1\,TeV and all the curves are compatible within error bars above $\sim$300\,GeV. Error bars increase with the energy because the event statistic decreases dramatically. These error bars are not independent as the data corresponding to a given energy threshold are included in the lower energy analysis. The only visible degradation is near the reconstruction-level energy threshold ($<$200\,GeV), where the sensitivity is 5-10\% worse. For Moon data taken with reduced HV, the sensitivity degradation lies between 15\% and 30\%. It seems to increase with the NSB level, although above 400\,GeV the three curves are compatible within statistical errors. This degradation is caused by a combination of a higher extracted-signal noise (see section~\ref{sec:analysis}) and a smaller effective area. The degradation is even clearer in the UV-pass filter data, where the sensitivity is 60-80\% worse than the standard one. Such a degradation is expected, especially due to the fact that the filters reject more than 50\% of the Cherenkov light. Besides, sensitivity could also be affected by a poorer reconstruction of the images, especially in the pixels that are partially obscured by the filter frame ribs. At the highest energies ($>$2\,TeV) sensitivity seems to improve. This could be expected for bright images, that are less affected
by noise, but higher statistics at those energies would be needed to
derive further conclusions.

\begin{figure}[t]
\includegraphics[width=\columnwidth]{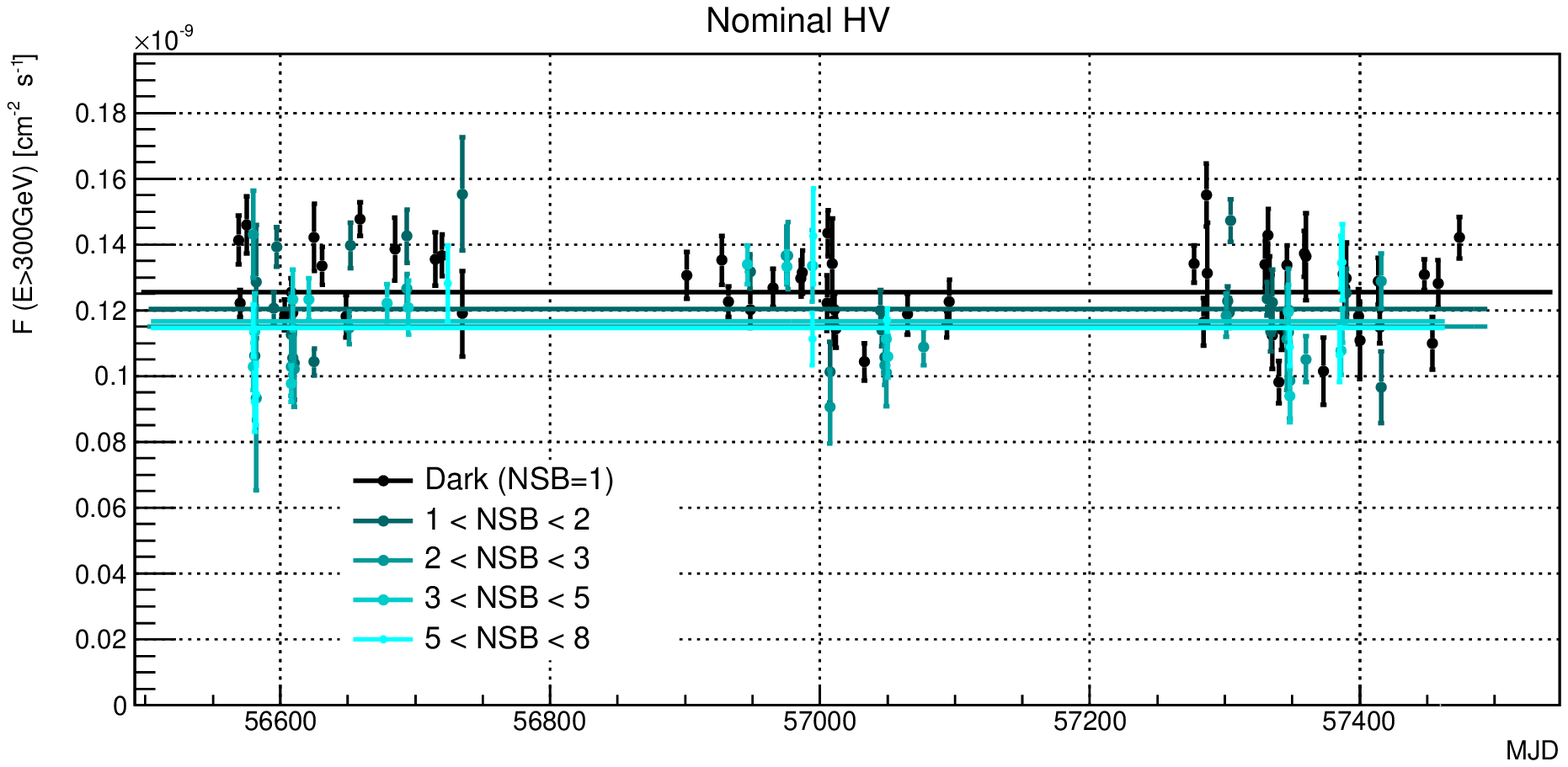}
\includegraphics[width=\columnwidth]{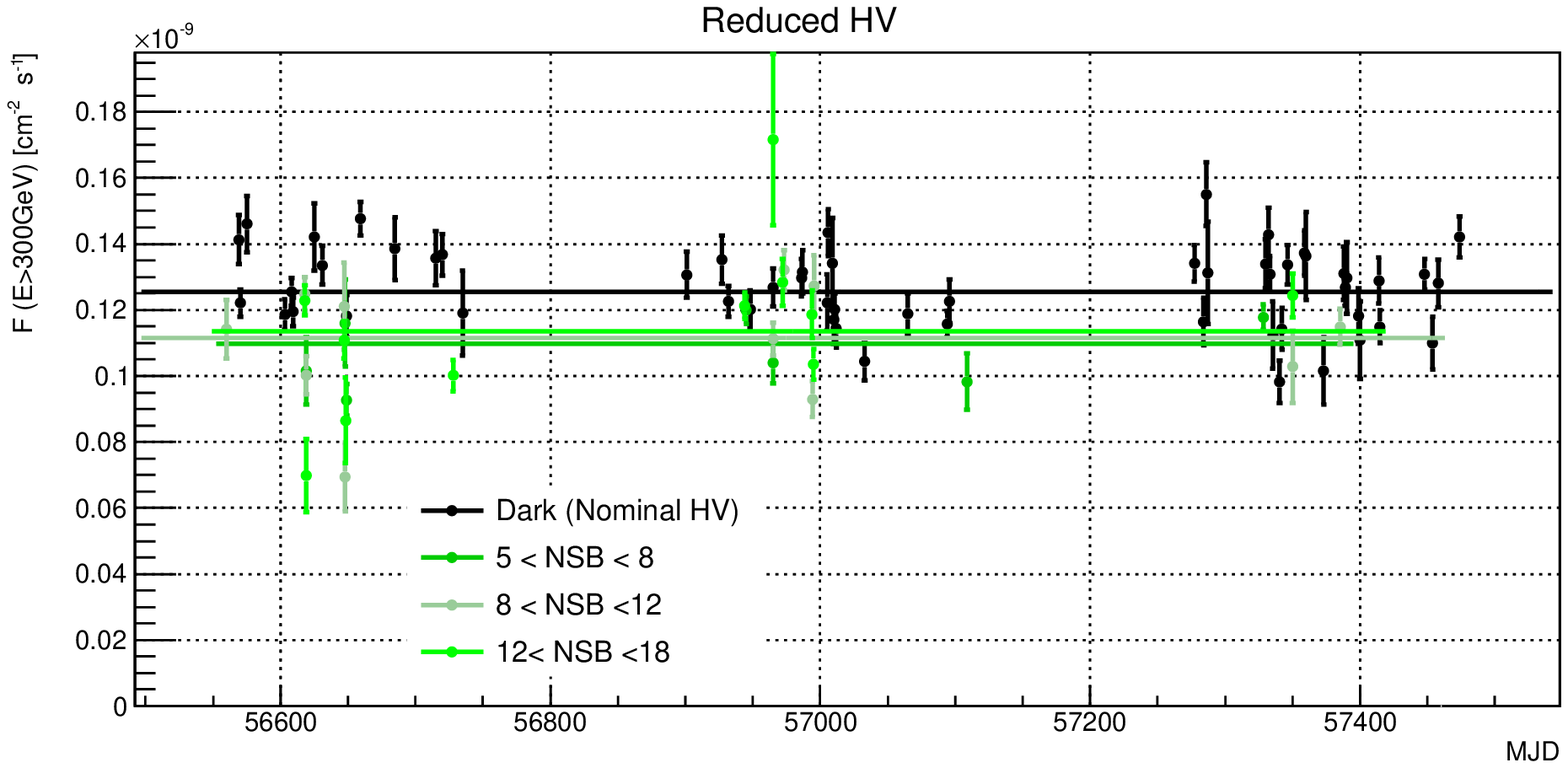}
\includegraphics[width=\columnwidth]{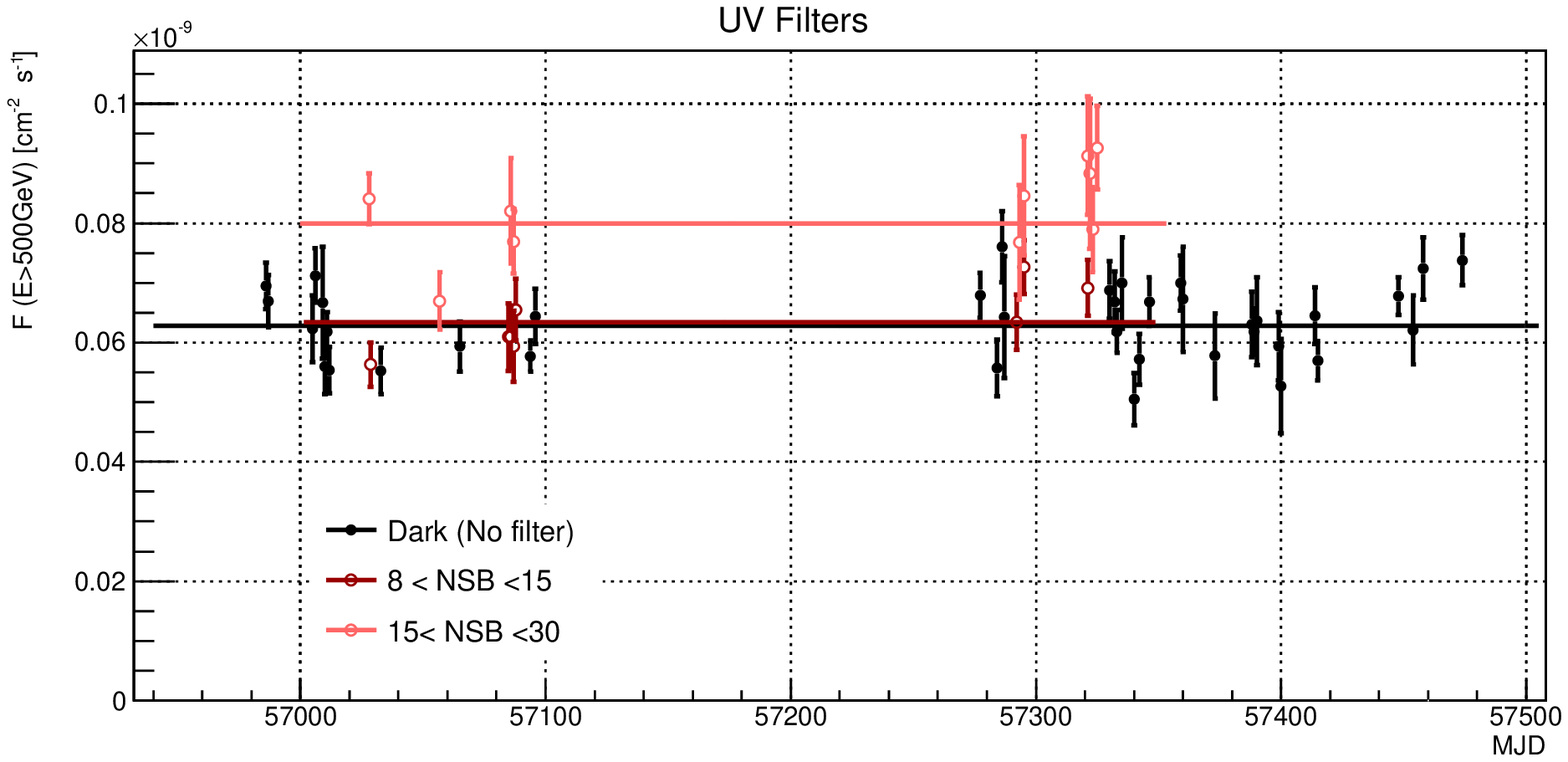}
\caption{Daily light curve of the Crab Nebula above 300\,GeV for observation under different sky brightness with nominal HV (top), reduced HV (middle) and above 500\,GeV for UV-pass filters (bottom). Horizontal lines correspond to the constant flux fit of the different NSB bins. For comparison, the LC and constant fit of the dark observation are reproduced in every panel.}\label{fig:LC}
\end{figure}

\subsection{Systematics}
During moonlight observations many instrumental parameters are more variable than during dark observations, in particular the trigger DTs and the extracted signal noise, and these variations induce larger MC/data mismatches and then larger systematic uncertainties. As shown in Section \ref{sec:CrabSpectra}, the Crab Nebula spectrum can be well reconstructed in every NSB/hardware bin. The reconstructed flux above the energy threshold of every NSB bin is within a 10\%, 15\%, 30\% error band around the flux obtained under dark conditions for nominal HV, reduced HV and UV-pass filter observations, respectively. The spectral shape is particularly well reproduced in all hardware configurations. The dark-Moon flux ratios vary less than 10\% over an order of magnitude in energy, corresponding to an additional systematic on the power-law spectral index below 0.05. 

The overall flux may mask large day-to-day fluctuations due to different sky brightness. To estimate this additional day-to-day systematic, we show in figure~\ref{fig:LC} the daily light curve (LC) of the Crab Nebula flux above 300\,GeV from October 2013 to March 2016 for every NSB level observed without UV-pass filters and the LC above 500\,GeV from January to October 2015 for the two NSB bins with UV-pass filters\footnote{UV-pass filter observation started only in January 2015. We use higher cut in energy for the UV-pass filter LC because the last bin (NSB:15-30$ \times  \, \textit{NSB}_{\text{Dark}}$) has an energy threshold above 300\,GeV at the observed zenith angles.}. Taking into account only statistical fluctuations, the $\chi^2$ test indicates that a constant flux is incompatible for every LC (even for dark observations). Assuming conservatively that the additional fluctuations are only due to systematic uncertainties (i.e., the Crab Nebula flux is constant), we estimate these systematic uncertainties by adding errors quadratically to the statistical errors in every data point until the constant-fit $\chi^2$ equals the number of degrees of freedom $k$ plus or minus $\sqrt{2k}$ (standard deviation of the $\chi^2$ distribution).
In order to constrain strongly the constant fit we include data points of several NSB bins for the fit of moderate moonlight with nominal HV ($1-8 \times  \, \textit{NSB}_{\text{Dark}}$), moonlight with reduced HV ($5-18  \times  \, \textit{NSB}_{\text{Dark}}$) and strong moonlight with UV-pass filter ($8-30  \times  \, \textit{NSB}_{\text{Dark}}$). Table~\ref{tabSyst} gives the day-to-day systematic errors obtained for these three hardware/NSB conditions as well as for dark observation with nominal HV.

 \begin{table*}[t]
\centering
\begin{tabular}{| c | c | c |}
\hline
 Sky Brightness & Hardware Settings & Day-to-day Systematics \\
\hline
     Dark ($\textit{NSB}_{\text{Dark}} = 1$) & nominal HV & $(7.6\pm1.2)$\%  \\
     1-8 $\textit{NSB}_{\text{Dark}}$ & nominal HV & $(9.6\pm1.2)$\% \\
     5-18 $\textit{NSB}_{\text{Dark}}$ & reduced HV & $(15.4\pm3.2)$\% \\
     8-30 $\textit{NSB}_{\text{Dark}}$ & UV-pass filters & $(13.2\pm3.4)$\% \\
   \hline
 \end{tabular}
 \caption{Additional systematic uncertainties that must be added to the errors of the LC shown in Figure \ref{fig:LC} to get constant-fit $\chi^2$ equaling the number of degrees of freedom. In the UV-pass filter case, the computed day-to-day systematic errors are valid for energies above 500 GeV.}
 \label{tabSyst}
 \end{table*}

For dark observations, the obtained day-to-day systematic uncertainty is $(7.6\pm1.2)$\%. This result is below the previous study based on Crab Nebula LC that reports a day-to-day systematic uncertainty of $\sim$12\% for the period from November 2009 to January 2011 \citep{Perf2012} and from October 2009 to April 2011 \citep{Crab2015JHEAp}. This is consistent with the result after the telescope upgrade reported in \citep{upgrade2}, which claims day-to-day systematic uncertainty below 11\%. For observation under moonlight with nominal HV (NSB~$<8 \times  \, \textit{NSB}_{\text{Dark}}$), the obtained day-to-day systematic is $(9.6\pm1.2)$\%, still below the 11\%. The additional systematic due to the moonlight is marginal and can be only constrained to be below 9\%. For brighter moonlight that requires hardware modifications, the systematic errors get larger. A few data points show a flux much lower than expected (down to $\sim$50\%). The overall day-to-day systematic is estimated at $(15.4\pm3.2)$\% for reduced HV and $(13.2\pm3.4)$\% for UV-pass filters, corresponding to an additional systematic on top of the dark nominal HV systematic errors laying between 6\% and 18\%.
For every hardware configuration, the additional day-to-day systematic errors is of the same order, or below, the systematic errors found for the overall flux.

To summarize, the additional systematic uncertainties of MAGIC during Moon time depend on the hardware configuration and the NSB level. For moderate moonlight (NSB~$<8 \times  \, \textit{NSB}_{\text{Dark}}$) observations with nominal HV, the additional systematic errors on the flux is below 10\%, raising the flux-normalization uncertainty (at a few hundred GeV) from 11\% \citep{upgrade2} to 15\%. For observations with reduced HV (NSB~$<18 \times  \, \textit{NSB}_{\text{Dark}}$) the additional systematic errors on the flux is $\sim$15\%, corresponding to a full flux-normalization uncertainty of 19\% after a quadratic addition. For UV-pass filter observations, the flux-normalization uncertainty increases to 30\%. The additional systematic on the reconstructed spectral index is negligible ($\pm$0.04) and the overall uncertainty is still $\pm$0.15 for all hardware/NSB configurations. The uncertainty of the energy scale is not affected by the moonlight. It may increase for reduced HV and UV-pass filter observations but this effect is included in the flux-normalization uncertainty increase\footnote{It is difficult to determine if a flux shift is due to wrong energy calibration or wrong effective area calculation.}. Concerning the pointing accuracy, as discussed in Section \ref{sec:AngRes}, no additional systematic uncertainties have been found.

\section{Conclusions}

For the first time the performance under moonlight of an IACT system is studied in detail with an analysis dedicated for such observations, including moonlight-adapted MC simulations. This study includes data taken with three different hardware settings: nominal HV, reduced HV and UV-pass filters.

During moonlight, the additional noise results in a higher energy threshold increasing with the NSB level, which for zenith angles below 30$^{\circ}$ goes from $\sim$70\,GeV (at the reconstruction level) under dark conditions up to $\sim$300\,GeV in the brightest scenario studied (15-30~$\times  \, \textit{NSB}_{\text{Dark}}$). With a dedicated moonlight-adapted analysis, we are able to reconstruct the Crab Nebula spectrum in all the NSB/hardware bins considered. The flux obtained is compatible within 10\%, 15\% and 30\% with the one obtained under dark conditions for nominal HV, reduced HV and UV-pass filter observations, respectively. The systematic uncertainty on the flux-normalization, 11\% for standard dark observation, increases to 15\% for nominal HV moonlight observations with NSB $< 8\times  \, \textit{NSB}_{\text{Dark}}$, 19\% for reduced HV observations between 5 and 18 $\times  \, \textit{NSB}_{\text{Dark}}$ and 30\% for UV-pass filter observations between 8 and 30 $\text{NSB}_{\text{Dark}}$. No significant additional systematic on the spectral slope was found, and the overall uncertainty is still $\pm$0.15 as reported in \cite{upgrade2}.

An eventual degradation in the sensitivity is constrained to be below 10\% while observing with nominal HV under illumination levels $<8 \times  \, \textit{NSB}_{\text{Dark}}$. The sensitivity degrades by 15 to 30\% when observing with reduced HV and by 60 to 80\% when observing with UV-pass filters. No significant worsening on the angular resolution above 300\,GeV was observed.

The main benefit of operating the telescopes under moonlight is that duty cycle can be doubled, suppressing the need to stop observations around full Moon. Depending on the needed energy threshold, many projects can profit from this additional time. Already moderate moonlight observations lead to the discovery of several active galactic nuclei, such as PKS~1222+21 \citep{PKS1222}, 1ES~1727+502 \citep{MAGIC1ES1727, Veritas1ES}, B3~2247+381 \citep{B322}. They are also used to study light curves of variable sources with better sampling, for instance the binary systems LSI~+61~303 \citep{LSI} and HESS~J0632+057 \citep{HESSj0632} and the active galactic nuclei PG1553+13 \citep{PG1553}, or to accumulate large amount of data as for deep observations of the Perseus cluster \citep{PerseusCR}.

The present study shows that, except for the energy threshold, the performance of IACT arrays is only moderately affected by moonlight. Hardware modifications to tolerate a strong sky brightness (reduced HV, UV-pass filters) seem to have more effect than the noise increase. The use of robust photodetectors, e.g. silicon photomultipliers, in the future should improve the performance under these bright conditions. The bright moonlight observations are particularly useful for projects in which the relevant physics lie above a few hundred GeV, such as long monitoring campaigns of VHE sources with hard spectrum or deep observation of supernova remnants for PeVatron studies. The eventual loss in sensitivity can be compensated with the possibility of much longer observation time in a less demanded observation period (currently often even used for technical work). In addition, observations under extreme NSB conditions are sometimes unavoidable, as in the case of the observation of the shadowing of cosmic rays by the Moon\footnote{Under such conditions the NSB level can be much higher than the 30~$\times  \, \textit{NSB}_{\text{Dark}}$ limit until which the performance was studied here.}. Observations under moonlight open many possibilities that should be more and more used with the current flourish of the VHE $\gamma$-ray astronomy using the IACT.

\section*{Acknowledgments}
%
%

We would like to thank the Instituto de Astrof\'{\i}sica de Canarias for the excellent working conditions at the Observatorio del Roque de los Muchachos in La Palma. The financial support of the German BMBF and MPG, the Italian INFN and INAF, the Swiss National Fund SNF, the ERDF under the Spanish MINECO (FPA2015-69818-P, FPA2012-36668, FPA2015-68378-P, FPA2015-69210-C6-2-R, FPA2015-69210-C6-4-R, FPA2015-69210-C6-6-R, AYA2015-71042-P, AYA2016-76012-C3-1-P, ESP2015-71662-C2-2-P, CSD2009-00064), and the Japanese JSPS and MEXT is gratefully acknowledged. This work was also supported by the Spanish Centro de Excelencia ``Severo Ochoa'' SEV-2012-0234 and SEV-2015-0548, and Unidad de Excelencia ``Mar\'{\i}a de Maeztu'' MDM-2014-0369, by the Croatian Science Foundation (HrZZ) Project 09/176 and the University of Rijeka Project 13.12.1.3.02, by the DFG Collaborative Research Centers SFB823/C4 and SFB876/C3, and by the Polish MNiSzW grant 2016/22/M/ST9/00382.

\section*{References}

\bibliography{mybibfile}

\end{document}